\documentclass[pra,amsmath,amssymb,twocolumn]{revtex4}
\usepackage{graphicx}
\usepackage{amsthm}

\newcommand{\diff}{\mathrm{d}}

\newcommand{\ket}[1]{\left | #1 \right \rangle}
\newcommand{\bra}[1]{\left \langle #1 \right |}
\newcommand{\braket}[2]{\left \langle #1 \, \right | \left. #2 \right \rangle}

\DeclareMathOperator{\trace}{tr}

\newcommand{\CLMZLowerlim}{\ensuremath{-\frac{\pi}{2}}}
\newcommand{\CLMZUpperlim}{\ensuremath{\frac{\pi}{2}}}

\begin{document}

\title{A measurement-based measure of the size of macroscopic quantum superpositions}
\author{Jan Ivar Korsbakken$^{1,2}$}
\author{K.\ Birgitta Whaley$^{1}$}
\author{Jonathan Dubois$^{1}$}
\affiliation{Berkeley Center for Quantum Information and Computation, Departments of Chemistry$^{1}$ and Physics$^{2}$, University of California, Berkeley, California 94720}
\author{J.\ Ignacio Cirac}
\affiliation{Max Planck Institut f\"{u}r Quantenoptik, Hans-Kopfermann Strasse 1, D-85748 Garching, Germany}

\begin{abstract}
Recent experiments claiming formation of quantum superposition states in near macroscopic systems raise the question of how the sizes of general quantum superposition states in an interacting system are to be quantified.  We propose here a measure of size for such superposition states that is based on what measurements can be performed to probe and distinguish the different branches of the state. The measure allows comparison of the effective size for superposition states in very different physical systems. It can be applied to a very general class of superposition states and reproduces known results for near-ideal cases. Comparison with a prior measure based on analysis of coherence between branches indicates that significantly smaller effective superposition sizes result from our measurement-based measure.  Application to a system of interacting bosons in a double-well trapping potential shows that the effective superposition size is strongly dependent on the relative magnitude of the barrier height and interparticle interaction.
\end{abstract}

\maketitle

\section{Introduction}
\label{sec:intro}

Despite quantum mechanics being one of the most sweepingly successful theoretical frameworks in the history of physics, there has always been and still appears to be a great deal of unease and confusion about some of its fundamental concepts and consequences. Most strikingly, quantum mechanics requires that if the outcomes of certain experiments are known with certainty, then it will not be possible to predict the outcome of other, incompatible experiments.
Instead, the system must exist in an indeterminate state, allowing for the possibility of several different outcomes of these experiments. In many interpretations, this is viewed as the system simultaneously existing in a ``superposition'' of all the different outcomes at once, until an experiment is actually performed and an outcome determined.

This seemingly ghostly state of affairs is perhaps not very unnerving in the context of atoms and microscopic systems. But, as Schr\"{o}dinger pointed out in 1935 \cite{Schrodinger1935}, a microscopic system coupled to a macroscopic one would inevitably lead to a situation in which even a macroscopic living being --- in his example a cat --- could conceivably end up in a state of being neither alive nor dead, until an observer actually looks and determines its fate. One ``solution'' proposed by some people uncomfortable with this situation, is that there may be some intrinsic ``size limit'' for quantum mechanics, which somehow prohibits nature from putting macroscopic systems into this kind of counter-intuitive superposition (see e.g., Ref.~\cite{Leggett2002} for a review). Although one may doubt such a proposal or question the need for it, it does deserve to be investigated whether it can be formulated in a precise enough way to be tested experimentally, especially given claims in recent years that ``Schr\"{o}dinger cat'' states have been or can be produced in more or less macroscopic systems \cite{WalHaarWilhelm2000,FriedmanPatelChen2000,JulsgaardKozhekinPolzik2001,MassarPolzik2003,MarshallSimonPenrose2003,Leggett2002}.

In order to investigate any possible size limits to quantum mechanics experimentally, one must of course have a reasonably clear definition of what the size of a system involved in quantum coherent behaviour is.  
In this paper we will investigate systems described by cat-like states that can be generically written as $\ket{\Psi} \propto \ket{A} + \ket{B}$, where $\ket{A}$ and $\ket{B}$ are macroscopic or mesoscopic states that are distinguishable to some extent. The task is to define a measure of how ``large'' this quantum superposition is in terms of the constituent subsystems.   Each of these notions will be made more precise in the course of this paper.  We explicitly seek a measure that is independent of the physical nature of the subsystems and that can therefore be used to compare the effective size of cat-like states realized in very different physical situations, e.g., Bose Einstein condensates and superconducting current loops.  

This question, which could be succinctly phrased as ``how big is Schr\"{o}dinger's cat'' for a given system in a particular quantum superposition state, has been asked in several earlier papers~\cite{Leggett2002,DurSimonCirac2002}.
By size we mean the number of effective independent subsystems that can describe the superposition (we will discuss in more detail what we mean by these notions in Section~\ref{sec:catsize}). One ``ideal'' $N$-particle cat state, for which the answer would be $N$, is a GHZ-state of the form $\ket{\Psi} = 2^{-1/2} \left ( \ket{0}^{\otimes N} + \ket{1}^{\otimes N} \right )$, where $\ket{0}$ and $\ket{1}$ are any pair of orthogonal one-particle states. Hardly any states realizable in the laboratory are of this idealized form however, and we therefore seek a measure that can quantify the size of more general states that are still recognizable as generic cat-like states but that may be very different from the ideal form.   The particular case of a generalized GHZ-like $N$-particle state of the form $\ket{\Psi} = K^{-1} \left ( \ket{\phi_0}^{\otimes N} + \ket{\phi_1}^{\otimes N} \right )$,
where $\ket{\phi_0}$ and $\ket{\phi_1}$ are non-orthogonal one-particle states, was studied in \cite{DurSimonCirac2002} with two independent approaches, one based on the stability with respect to decoherence, and the other on the amount of distillable entanglement.  
In the limit of highly overlapping states, where $\left | \braket{\phi_0}{\phi_1} \right |^2 = 1 - \epsilon^2$ for some $\epsilon \ll 1$, D{\" u}r et al. found that 
both decoherence and distillable entanglement measures of the ``effective'' number of degrees of 
freedom participating in the superposition, $n$, yielded an effective cat size of $n \sim N\epsilon^2$ \cite{DurSimonCirac2002}. These two measures were specific to the form of the non-orthogonal GHZ-like states and it is not obvious how to apply them to arbitrary superposition states.  Another motivation of this paper is thus to derive a measure of the effective cat size that can be applied to superpositions $\ket{A} + \ket{B}$ having completely general forms of the states $\ket{A}$ and  $\ket{B}$.   

The rest of this paper is divided into four parts. In section~2, we present a measure of effective ``cat size'' for general binary superposition states that is based on the notion that the ``cattiness'' of a superposition state should depend primarily on how \textit{distinguishable} the two branches of the state are.  This measure is based fundamentally on measurements, and thereby differs from earlier measures that have tended to be based on the mathematical form of the state. The new measure is thus potentially more useful for experimental implementations.  In section~3, we apply the measure to a system of bosons in a two-mode description. In section~4 we connect the results from section~2 with realistic numerical Monte Carlo simulations of Bosons with attractive interactions trapped in a double-well potential. Section~5 summarizes and indicates future directions of research.

\section{Ideal cats and effective cat sizes}
\label{sec:catsize}

In this Section we will give a definition of the size of a
cat-like state of an object, $\ket{\Psi} = \ket{A} +
\ket{B}$. We will consider that the object is formed by $N$
subsystems, and our measure of effective size will then range between $0$ and $N$,
analogously as in Ref. \cite{DurSimonCirac2002}. However, in contrast to that
work, the quantity we introduce here will measure how
(macroscopically)  distinguishable the states $\ket{A}$ and
$\ket{B}$ are. The main idea that we want to capture with this
definition is the following: how many fundamental subsystems of the object do we have to measure in order to
collapse the entire state into a single branch corresponding to one
of the two states $\ket{A}$ or $\ket{B}$, and how many times larger than this number is the entire system?  By ``fundamental subsystem'', we mean something that in some sense can be taken as a fundamental building block of our system, e.g., single particles or something similar.
It is by no means always clear what one should consider the fundamental building blocks of a given physical system (molecules, atoms, Cooper pairs, electrons, quarks...), and we will not attempt to make a definitive definition of what such building blocks should be, if this is even possible. However, our measure will be based on how many measurements must be carried out to perform a specific task, namely to collapse the superposition state into one branch or the other. A reasonable qualitative definition would therefore be that a fundamental subsystem is the smallest subsystem that one could in principle measure in some experimental context and which would provide information that could help distinguish one branch from the other. For a BEC experiment one could in principle, e.g., scatter light from single atoms, making single atoms reasonable candidates for fundamental subsystems.
Our measure thus will depend on the experimental situation and the relevant size and energy scale, something which probably must be expected if one wishes a measure that does not involve Planck-scale physics.
For the remainder of this paper, even though relevant fundamental subsystems may not always be something that can reasonably be called particles, we will use the terms ``particle'' and ``fundamental subsystem'' interchangeably, and this concept plays an important role in our measure. More specifically, the question we ask to define our measure is the
following: What is the maximal number of disjoint subsets that one can constitute from the
$N$ particles such that by measuring all particles in any given subset
one can cause the superposition state to collapse into one of the branches $\ket{A}$ or $\ket{B}$ to a high degree. A measurement that causes such a collapse is equivalent to a measurement that with high probability lets us determine correctly whether a system \textit{is} in state $\ket{A}$ or $\ket{B}$ if we are given a system which is definitely in either one of these two states, but we do not know which one. We emphasize that the latter situation is clearly very different from having a system which is actually in a superposition $\ket{A}+\ket{B}$. But since a measurement which collapses the superposition state is idential to one which is capable of distinguishing between the two branch states (assuming an ideal measurement with no classical noise), we shall often use the latter picture in the discussion below

It is not difficult to write a mathematical definition which expresses our measure as formulated above. However, in practice it may
be quite difficult to calculate this for general superpositions, since for a given accuracy one has to optimize the number of subsets over all possible
partitionings of the $N$ particles.  Thus, we will use an alternative definition that also
captures the above concepts but is simpler to evaluate, particularly for states possessing permutation invariance.

\theoremstyle{plain}
\newtheorem*{catsizedef}{Definition of cat size}
\begin{catsizedef}
Given  an object composed of $N$ subsystems and $0<\delta\ll 1$, we define the cat size of a state $\ket{\Psi} \propto  \ket{A} + \ket{B}$ with $||\ket{A}|| = ||\ket{B}|| = 1$, to a precision $\delta$, by
 \begin{equation}
 C_\delta(\Psi):=N/n_{\rm min},
 \label{eq:catsize}
 \end{equation}
where  $n_{\rm min}$ is the minimum number of particles one has to
measure, on average, in order to distinguish the 
states $\ket{A}$
and $\ket{B}$ with probability greater than or equal to 1-$\delta$.
\end{catsizedef}

In order to determine $C_\delta(\Psi)$ we can proceed as
follows. We begin with 1-particle measurements ($n=1$). For each particle $k$ we calculate the optimal
probability of being able to distinguish $\ket{A}$ and $\ket{B}$ by measuring just this particle 
and average this probability with
respect to $k$.  If the resulting average probability is larger than
$1-\delta$, then $n_{\text{min}}=1$ and hence $C_\delta(\Psi)=N$. If
not, we then go on to consider all possible sets of two particles, $(j,k)$, determining the
corresponding optimal probability of distinguishing $\ket{A}$ and $\ket{B}$ by measuring these two particles. If after averaging
this probability with respect to $j,k$ we obtain an average probability larger than $1-\delta$,
we have $n_{\text{min}}=2$ and hence $C_\delta(\Psi)=N/2$. If not,
we repeat the procedure with measurements of an increasing number of particles until we
reach a value of $n_{\text{min}}$ for which the averaged probability of successfully distinguishing the two branches is for the first time larger than $1-\delta$.
If this happens only when all particles are measured, then $n_{\text{min}} = N$, and the cat size is $C_{\delta} = 1$. If even measuring all $N$ particles still fails to distinguish the two branches to the desired precision $1-\delta$ , then $n_{\text{min}}$ and hence the cat size $C_{\delta}$ are essentially undefined. For simplicity, we will define the cat size to be zero in this situation.

Thus, the only ingredient we need in order to determine the cat size is the maximal probability to be able to distinguish two states $\ket{A}$ and $\ket{B}$ by measuring only a given subset of the total system (or using some similarly restricted set of measurements, as we will see in Section~\ref{sec:bosonic}). We now briefly discuss this probability. For more thorough and general discussions, see Refs.~\cite{Helstrom1976,Fuchs1995,FuchsGraaf1999}.
Using a generalized quantum measurement, {\it i.e.}, a POVM (positive operator valued measure)~\cite{Peres1995}, in which the outcome described by POVM element $E_A$ is taken to indicate that the system is in state $\ket{A}$ and the outcome $E_B$ that it is in $\ket{B}$, then given equal prior probabilities for each state (i.e. equal weight for the two branches of the superposition), the probability $P$ of inferring the correct state from a single measurement is
\begin{equation}
P = \frac{1}{2} \left [ \trace \left ( \rho_A E_A \right ) + \trace \left ( \rho_B E_B \right ) \right ],
\end{equation}
where $\rho_A = \ket{A}\bra{A}$ and $\rho_B = \ket{B}\bra{B}$ are the density matrices of the two states. If we now restrict ourselves to measure only a subset of $n$ particles, then the measurement outcomes are given by POVM elements $E^{(n)}_{A}, E^{(n)}_{B}$ that act non-trivially only on these $n$ particles,
acting as the identity on the remaining $N-n$ particles. The probability of successfully inferring the state is then\begin{equation}
\begin{split}
P &= \frac{1}{2} \left [ \trace \left ( \rho_A \, E_A^{(n)} \otimes \openone^{(N-n)} \right ) + \trace \left ( \rho_B \, E_B^{(n)} \otimes \openone^{(N-n)} \right ) \right ] \\
&= \frac{1}{2} \left [ \trace \left ( \rho_A^{(n)} E_A^{(n)} \right ) + \trace \left ( \rho_B^{(n)} E_B^{(n)} \right ) \right ],
 \end{split}
 \label{eq:SuccessProb}
\end{equation}
with $\rho_{A}^{(n)} \equiv \trace_{N-n} \rho_{A}$ and $\rho_{B}^{(n)} \equiv \trace_{N-n} \rho_{B}$ the corresponding n-particle reduced density matrices ($n$-RDMs).  ($\trace_{N-n}$ denotes the trace over all particles except the $n$ particles being measured.)  The maximum probability for successfully distinguishing two density matrices $\rho_A^{(n)}$ and $\rho_B^{(n)}$ will then be given by an optimal POVM, which is known to be a projective measurement in the eigenbasis of the operator $\rho_A^{(n)} - \rho_B^{(n)}$:~\cite{Helstrom1976, FuchsGraaf1999, Fuchs1995}
\begin{equation}
P = \frac{1}{2} + \frac{1}{4} || \rho_A^{(n)} - \rho_B^{(n)} ||.
\label{eq:PE_RDM}
\end{equation}
Here $||X||={\rm tr}|X|$ is the trace norm, i.e., $ \sum_i |\lambda_i|$, with $\lambda_i$ the eigenvalues of $X$.

\renewcommand{\labelenumi}{(\roman{enumi})}
Several remarks are in order here.
\begin{enumerate}
\item We have based our working definition here
on the average probability over all equal size subsets being larger than $1-\delta$. One could alternatively have employed a requirement that the
minimal probability is larger than 1-$\delta$.  Also, as mentioned above, at the cost of introducing a great deal more computational expense, one could replace the average over equal size subsets by the optimum partition over all possible subsets.
\item Although it should be clear from the notation, we note that, as defined, our measure applies only to pure quantum states, not mixed states. Defining a cat size measure for mixed states is complicated by the fact that there is no unique way to decompose a mixed state density matrix into a convex sum of pure states, so that, e.g., a mixture of completely separable states could also be written formally as a mixture of very cat-like states. Any cat size measure applicable to mixed states would therefore have to weight the cat size quite heavily with the purity of the state. We will not pursue such an extension of our measure in this paper.
\item For states
that are symmetric with respect to permutations, for a given
number of measured particles $n$ it suffices to consider only a single
subset, since all subsets give rise to the same probability because of symmetry. This
results in a considerable gain for computational studies with large $N$ and will be analyzed in detail for bosonic systems in the remainder of this paper.  
\item We have assumed
that we can perform collective measurements on a subset of $n$
particles.  However, we can also consider the situation in which only individual single-particle
measurements are performed. In some cases the calculation could then be
highly simplified, since we would have to consider only single-particle
reduced density operators. This situation appears well suited to 
bosonic systems and will be analyzed further in Section~\ref{subsec:bosonic_mx1}.
\item Given a state $\Psi$ in which $\ket{A}$
and $\ket{B}$ are not specified, there are many ways of selecting the two branches,
and these may give rise to different values of the measure. Thus, when we talk
about the size of a cat state, we must always specify what are the branches $A$ and $B$.  Furthermore, application of the measurement-based cat size defined above requires that the two branches have the same norm.
If the norm of the two branches are different, i.e., 
\begin{equation}
\ket{\Psi} \propto \ket{A} + g \ket{B},
\label{eq:UnbalancedCat}
\end{equation}
with $0 \leq g \leq  1$,
we expect that $C_\delta(\Psi)$ must be multiplied by a factor that interpolates smoothly between a value of zero when $g=0$ and a value of unity when $|g|=1$.  This factor can be determined by recognizing that the general superposition for general $g$ can always be distilled to the equal superposition $|g|=1$ by generalized measurements~\cite{DurSimonCirac2002}, yielding an effective cat size that is reduced by the associated probability. For the state 
of Eq.~(\ref{eq:UnbalancedCat}), one can perform a measurement using the operator
\begin{equation}
A_1 \equiv \frac{g \ket{A} \bra{B_{\perp}}}{\braket{B_{\perp}}{A}} + \frac{\ket{B}\bra{A_{\perp}}}{\braket{A_{\perp}}{B}}
\label{eq:CatBalancingMeasurement}
\end{equation}
and complement this with any other measurement operator $A_2$ such that $E_1 \equiv A_1^{\dag} A_1$ and $E_2 \equiv A_2^{\dag} A_2$ form a POVM, 
{\it i.e.}, $E_1 + E_2 = \openone$. $\ket{A_{\perp}}$ and $\ket{B_{\perp}}$ are any states that are orthogonal to $\ket{A}$ and $\ket{B}$ respectively. If one obtains the outcome corresponding to $A_1$, then the state after the measurement will be the equal superposition state $\ket{\Psi} \propto \ket{A} + \ket{B}$. The probability for this to happen is
\begin{equation}
p_g = \frac{\left ( 2 + \braket{A}{B} + \braket{B}{A} \right ) |g|^2}{1 + \braket{A}{B} g + \braket{B}{A} g^* + |g|^2}
\label{eq:CatBalancingProbability}
\end{equation}
Thus if the norm of the two branches are different, we can take the effective cat size to be $p_g C_\delta (\Psi)$, where $g$ is the smaller of the two norms. Note that if at least one particle separates out in each of the branches $\ket{A}$ and $\ket{B}$, {\it i.e.}, if $\ket{\Psi}$ can be written in the form $\ket{a}\ket{A^{N-1}} + \ket{b}\ket{B^{N-1}}$ for some one-particle states $\ket{a}$, $\ket{b}$ and $(N-1)$-particle states $\ket{A^{N-1}}$, $\ket{B^{N-1}}$, then the distillation can be accomplished using only a local single-particle measurement, namely
\begin{equation}
A_1 = \frac{g\ket{a}\bra{b}}{\braket{b_{\perp}}{a}} + \frac{\ket{b}\bra{a_{\perp}}}{\braket{a_{\perp}}{b}}
\end{equation}
The probability of obtaining the outcome $A_1$ is the same as in Eq.~(\ref{eq:CatBalancingProbability}),
with $\ket{A}$ replaced by  $\ket{a}\ket{A^{N-1}}$.

\item  In order to calculate $C_{\delta}(\Psi)$, we can calculate $P=1-P_E$, where $P_E$ is the probability of error in distinguishing the two states, and then find the value of $n$ for which $P > 1- \delta$. In Section~\ref{sec:bosonic} we will show plots of $P_E$ rather than $P$, since these better illustrate
the scaling of the error with $n$.  For large $N$ values, in some situations we can also solve $P = 1-\delta$ to obtain a continuous value of $n$ (see Section~\ref{subsec:bosonic_mx1}).

\item Our approach of asking how many subsystems a system can be divided into such that each one alone suffices to distinguish the branches of a state, has some similarities with the concept of \textit{redundancy}, introduced in a different context in~\cite{Blume-KohoutZurek2006}. There, the redundancy of a piece of information about a quantum system is defined as the number of fragments (partitions, in our terminology) into which the environment can be divided such that this information is contained in every one of the fragments. This is used in~\cite{Blume-KohoutZurek2006} to probe how \textit{objective} a certain piece of information about a quantum system is, since information that has a high degree of redundancy can be obtained by many observers independently through measuring different parts of the environment, without disturbing the system itself or each other's measurements.

\item Finally, we note that our measure does not look at the physical properties of
the object, such as mass or spatial dimensions, but rather at the number of components.
For example, with this measure a very massive elementary particle can have a
cat size of 1 at most.

\end{enumerate}

We now give some examples of the cat size for simple superposition states, calculated using the above formalism in a two-state basis. Suppose
we have a system consisting of a macroscopic number $N$ of
spin-1/2 particles. First, consider the ideal GHZ states $\ket{\Psi_\pm} := (\ket{0}^{\otimes N} \pm \ket{1}^{\otimes N})$. Here only one particle need be measured to distinguish the two branches with certainty, i.e., the one-particle reduced density matrix (1-RDM) already gives $P=1$, and hence $n_{\text{min}}=1$ and $C_\delta=N$ for all $\delta$.
Now consider the linear superposition state $\ket{\Psi} = \frac{1}{\sqrt{2}} \left
( \ket{\Psi_{+}} + \ket{\Psi_{-}} \right ) = \ket{0}^{\otimes N} $.  This is also a superposition of two distinguishable (orthogonal) macroscopic quantum states, but here all $N$ particles must be measured in order to distinguish the two branches. The $n$-RDMs for $\ket{\Psi_+}$ and $\ket{\Psi_-}$ are identical for all $n<N$, so $P=0$ unless $n=N$, in which case $P=1$. Hence $n_{\text{min}} = N$ and the cat size is equal to 1, as expected since the state is equivalent to a product state.
As a final example, we apply our measure to the non-ideal state with non-orthogonal branches that was studied in Ref.~\cite{DurSimonCirac2002}, namely
$\ket{\Psi} := (\ket{0}^{\otimes N} + \ket{\phi}^{\otimes N})$ with $|\braket{0}{\phi}|^2=1-\epsilon^2$, where $\epsilon\ll 1$. Here, the two branches $\ket{0}^{\otimes N}$ and $\ket{\phi}^{\otimes N}$ are separable states, and their respective $n$-RDMs are therefore equal to density matrices of pure $n$-particle states, namely $\ket{0}^{\otimes n}$ and $\ket{\phi}^{\otimes n}$ respectively. In general, for any quantum system and any pair of states $\ket{a}$ and $\ket{b}$ with $|\braket{a}{b}|^2 = c^2$, we can write the corresponding density matrices in a two-state partial basis defined by $\ket{a}$ and $\ket{a_\perp}$, where $\ket{a_{\perp}}$ is the state orthogonal to $\ket{a}$ but contained in the subspace spanned by $\ket{a}$ and $\ket{b}$ \footnote{This basis does not necessarily span the whole Hilbert space, but we here only need the subspace spanned by $\ket{a}$ and $\ket{b}$.}. Specifically, writing $\ket{b} = c \ket{a} + s \ket{a_\perp}$ with $|c|^2 + |s|^2 = 1$, we have
\begin{eqnarray}
\rho_a - \rho_b &= &
     \left[
   \begin{matrix}
     1-|c|^2&-sc\\
     -sc& -|s|^2\\
    \end{matrix}
    \right]
\end{eqnarray}
Using Eq.~(\ref{eq:PE_RDM}), we find that $\ket{a}$ and $\ket{b}$ can be successfully distinguished with probability $P = \frac{1}{2} \left( 1 + |s| \right)$.
Defining $\ket{a} =\ket{0}^{\otimes n}$ and $\ket{b} = \ket{\phi}^{\otimes n}$, we then obtain the maximum success probability 
\begin{equation}
P = \frac{1}{2} \left( 1 + \sqrt{1-(1-\epsilon^2)^n} \right)
\label{eq:SuccessProbability}
\end{equation}
for distinguishing $\ket{0}^{\otimes N}$ and $\ket{\phi}^{\otimes N}$ using $n$-particle measurements.   Requiring this to be greater than $1-\delta$, where $\delta$ is the desired precision, results in a value of $n_\text{min}$ given by
\begin{equation}
n_\text{min} = \left \lceil \frac{\log(4\delta - 4\delta^2)}{\log(1-\epsilon^2)} \right \rceil
\end{equation}
where $\lceil \dots \rceil$ denotes the ceiling function, i.e., the nearest integer above the value of the argument.  For $\epsilon$ and $\delta$ small, this results in 
$C_\delta=N/n_\text{min} = N\epsilon^2/(-\log(\delta))$, in agreement with the $N\epsilon^2$ scaling found 
for these states in Ref.~\cite{DurSimonCirac2002}.

\section{Cat states in bosonic systems}
\label{sec:bosonic}

Most experiments involving quantum coherence in more or less macroscopic systems, including potentially macroscopic cat states, are performed on systems of identical particles. These include photon states~\cite{Brune1996}, superconducting current loops\cite{WalHaarWilhelm2000,FriedmanPatelChen2000}, 
spin-polarized atomic ensembles~\cite{JulsgaardKozhekinPolzik2001} and Bose Einstein Condensates~\cite{Donley2002}. 
Cat states of bosonic particles allow some simplification of the proposed measure of effective cat size, since making use of the permutation symmetry reduces the size and number of the $n$-RDMs to be analyzed.  
We consider here a generic form of cat state wavefunction that generalizes the ideal GHZ state
\begin{equation}
\ket{GHZ_N} = \frac{1}{\sqrt 2} \left( \ket{0}^{\otimes N} + \ket{1}^{\otimes N} \right)
\end{equation}
to situations described by a superposition of non-ideal GHZ-like states in which the single particle states are non-orthogonal.  In particular, we consider states of the form~\cite{Salgueiro2004} 

\begin{widetext}
\begin{equation}
\begin{split}
\ket{\Psi} &\propto \int_{\CLMZLowerlim}^{\CLMZUpperlim} \diff \theta \, f(\theta) \left [ \left ( \cos \theta \, a^{\dag} + \sin \theta \, b^{\dag} \right )^N + \left ( \sin \theta \, a^{\dag} + \cos \theta \, b^{\dag} \right )^N \right ] \ket{0} \\
&\equiv \int_{\CLMZLowerlim}^{\CLMZUpperlim} \diff \theta \, f(\theta) \left ( \ket{\phi^{(N)}_A (\theta)} + \ket{\phi^{(N)}_B (\theta)} \right ) \\
&\equiv  \left ( \ket{\Psi^{(N)}_A} + \ket{\Psi^{(N)}_B} \right ),
\label{eq:BECIntegratedCat}
\end{split}
\end{equation}
\end{widetext}

\noindent where the operators $a^{\dag}$ and $b^{\dag}$ create two orthogonal single-particle states. 
For fixed values of $\theta$, the integrand of Eq.~(\ref{eq:BECIntegratedCat}) corresponds to ground states of a two-state BEC with attractive interactions, found in~\cite{CiracLewensteinMolmer1998} using a two-mode approximation and an extended mean-field calculation.
In this section we will illustrate the effects of $f(\theta)$ for various values of its mean and variance.
The two branches of the superposition  $\ket{\Psi^{(N)}_A}$ and $\ket{\Psi^{(N)}_B}$ are thus defined here by a superposition of states $\ket{\phi^{(N)}_A (\theta)}$ and $ \ket{\phi^{(N)}_B(\theta)}$ that are themselves non-ideal GHZ-like states of variable orthogonality defined by the angle $\theta$. In the notation above, $\theta=0$ 
and $\pi/2$ correspond to perfect orthogonality of the single-particle states $\ket{\phi^{(1)}_A (\theta)} = (\cos \theta \, a^{\dag} + \sin \theta \, b^{\dag})\ket{0}$ and $ \ket{\phi^{(1)}_B (\theta)} = (\sin \theta \, a^{\dag} + \cos \theta \, b^{\dag})\ket{0}$ (with $\ket{\phi^{(1)}_A(0)} = a^{\dag}\ket{0}$ and $\ket{\phi^{(1)}_B(0)} = b^{\dag}\ket{0}$, switched for $\theta = \pi/2$), $\theta=\pi/4$ corresponds to complete overlap (with both $\ket{\phi^{(1)}_A(\pi/4)}$ and $\ket{\phi^{(1)}_B(\pi/4)}$ equal to $2^{-1/2} ( a^{\dag} + b^{\dag})\ket{0}$), $\theta=-\pi/4$ also corresponds to complete overlap but with differing overall sign ($\ket{\phi^{(1)}_A(-\pi/4)} = - \ket{\phi^{(1)}_B(-\pi/4)} = 2^{-1/2} ( a^{\dag} - b^{\dag} )\ket{0}$), and $\theta= -\pi/2$ corresponds to orthogonality again but with a factor of $-1$ for each of the states relative to $\theta = \pi/2$. The extent to which the two branches can be delineated is clearly dependent on the amplitude function $f(\theta)$ that controls the amount of spreading of each branch.  The form of the spreading function $f(\theta)$  will depend on the details of the physical realization of the macroscopic superposition, as will the values of the angle $\theta$.
This generalized superposition reduces to the form employed in Ref.~\cite{CiracLewensteinMolmer1998} when $f(\theta)=\delta (\theta-\theta_0)$ for some $\theta_0$ dependent on the parameters of the Hamiltonian used there, and is in agreement with general expectations for the form of macroscopic superposition wave functions for superconductors \cite{Leggett2002}.
In Section~\ref{sec:DoubleWellMonteCarlo} we analyze the form of $f(\theta)$ appropriate to a cat state formed from a BEC trapped in an external double well potential.
Numerical calculations for attractive Bose gases have shown that the competing effects of tunneling between modes and interactions between particles can be taken into account by letting $f(\theta)$ be a Gaussian, the shape of which is determined by the ratio of tunnelling and interaction energies \cite{MuellerHoUeda2006, HoCiobanu2004}. Note that while Eq.~(\ref{eq:BECIntegratedCat}) is implicitly a two-mode wave function, this form can readily be generalized to multi-mode superpositions. 

\subsection{Calculation of effective cat sizes for superpositions of non-ideal states}
\label{subsec:bosonic_effectivecatsizes}

We can first give some qualitative expectations for the effective size of this superposition state when $N$ becomes large. There are two factors that will reduce the effective size below that of the ideal GHZ state, $N$.   Firstly, for values of $\theta \neq 0$,  the two branches of $\ket{\Psi}$ in Eq.~(\ref{eq:BECIntegratedCat}) are not orthogonal, and hence not completely distinguishable. As shown explicitly in Section~\ref{sec:catsize} above, our measure therefore gives a cat-size for this state that is smaller than $N$, in agreement with the results derived previously in Ref.~\cite{DurSimonCirac2002}. Second, if the amplitude function $f(\theta)$ deviates from a $\delta$-function, the inner product between the two branches will not approach zero even in the limit $N \rightarrow \infty$. Hence there will always be a finite minimal probability that we will not be able to distinguish the two branches, even in the thermodynamic limit and even if all $N$ particles are measured.  Eventually, if this irreducible overlap between the branches is large enough, the division into two different branches becomes meaningless. The effect of this second factor has not been investigated before, but is essential to investigate for understanding macroscopic superpositions in realistic physical systems.

To make quantitative calculations for states of the form of (\ref{eq:BECIntegratedCat}), it is convenient to 
first make a change of basis as follows,
\begin{align}
c &= \frac{1}{\sqrt{2}} \left ( a + ib \right ) & d &= \frac{1}{\sqrt{2}} \left ( b + ia \right ) \\
c^{\dag} &= \frac{1}{\sqrt{2}} \left ( a^{\dag} - i b^{\dag} \right ) &
d^{\dag} &= \frac{1}{\sqrt{2}} \left ( b^{\dag} - i a^{\dag} \right ),
\end{align}
so that the integrand components of the  two branches in Eq.~(\ref{eq:BECIntegratedCat}) become
\begin{align}
\ket{\phi^{(N)}_A(\theta)} &= \frac{1}{\sqrt{N!}2^{N/2}} \left ( e^{i\theta} c^{\dag} + ie^{-i\theta} d^{\dag} \right )^N \ket{0}, \nonumber \\
\ket{\phi^{(N)}_B (\theta)} &= \frac{1}{\sqrt{N!}2^{N/2}} \left ( e^{i\theta} d^{\dag} + ie^{-i\theta} c^{\dag} \right )^N \ket{0}.
\label{eq:newbasis}
\end{align}

When measuring indistinguishable bosons, we obviously cannot pick out $n$ specific particles to make an $n$-particle measurement as described in the discussion 
in Section~\ref{sec:catsize}. For indistinguishable particles, the Kraus operators \cite{KrausBohmDollard1983} describing the effect of any measurement outcome have the form, e.g., $A^{(n)}_k = \sum_{\{i\}} c_k^{i_1 i_2 \cdots i_n} a_{i_1} a_{i_2} \cdots a_{i_n}$, where $i$ denotes a single-particle state, with corresponding POVM elements
\begin{equation}
\begin{split}
E^{(n)}_k &= \sum_{\{i\}, \{j\}} (c_k^{i_1 \cdots i_n})^* \, c_k^{j_1\cdots j_n} \, a_{i_n}^{\dag} \cdots a_{i_1}^{\dag} a_{j_1} \cdots a_{j_n} \\
&\equiv \sum_{\{i\}, \{j\}} \left ( E^{(n)}_k \right )^{i_1 \cdots i_n}_{j_1 \cdots j_n} \, a_{i_n}^{\dag} \cdots a_{i_1}^{\dag} a_{j_1} \cdots a_{j_n} .
\end{split}
\label{eq:BosonicPOVM}
\end{equation}
Here $k$ labels the outcome and the superscript $(n)$ specifies the number of particles on which the operator acts \footnote{POVM elements obtained from Kraus operators containing both creation and annihilation operators can also be expressed as a linear combination of products of some number of creation operators followed by the same number of annihilation operators, except that some terms may contain less than $2n$ operators.}.

Eq.~(\ref{eq:BosonicPOVM}) gives us the probability 
\begin{equation}
\begin{split}
P_k &= \trace \left ( \ket{\Psi}{\bra{\Psi}} \, E^{(n)}_k \right ) \\
&= \sum_{\{i\}, \{j\}} \left ( E^{(n)}_k \right )^{i_1 \cdots i_n}_{j_1 \cdots j_n} \bra{\Psi}  a_{i_n}^{\dag} \cdots a_{i_1}^{\dag} a_{j_1} \cdots a_{j_n} \ket{\Psi} \\
&\equiv \trace \left ( \mathcal{E}^{(n)}_k \rho^{(n)} \right ),
\end{split}
\end{equation}
for a given outcome $E^{(n)}_k$ when the system is in state $\ket{\Psi}$.
Here $\mathcal{E}^{(n)}_k$ is the matrix given by the coefficients $\left ( E^{(n)}_k \right )^{i_1 \cdots i_n}_{j_1\cdots j_n} \times N!/(N-n)!$, and
\begin{multline}
\left ( \rho^{(n)} \right )^{i_1 i_2 \cdots i_n}_{j_1 j_2 \cdots j_n} \equiv \frac{(N-n)!}{N!} \\ \times \bra{\Psi} a_{i_n}^{\dag} \cdots a_{i_2}^{\dag} a_{i_1}^{\dag} a_{j_1} a_{j_2} \cdots a_{j_n} \ket{\Psi}
\label{eq:BosonicRDM}
\end{multline}
is the $n$-particle reduced density matrix, or $n$-RDM, of the bosonic system in second quantized form. The combinatorial factors here are introduced so that $\rho^{(n)}$ will have trace $1$.
Furthermore, since $\rho^{(n)}$ is symmetric in both all upper and all lower indices, we can index $\rho^{(n)}$ by mode occupation numbers $k$ and $l$.
The resulting symmetrized matrix acts on a vector space which is equal to the 
full vector space projected onto a  symmetric subspace \cite{StocktonGeremiaDoherty2003}.
Denoting the symmetrized RDM by $\tilde{\rho}^{(n)}$, we obtain:
\begin{equation}
\left ( \tilde{\rho}^{(n)} \right )^k_l = \sqrt{{n\choose k}{n\choose l}} \, \left ( \rho^{(n)} \right )^{i_1 \cdots i_n}_{j_1 \cdots j_n}
\label{eq:SymmetrizedBosonicRDM}
\end{equation}
where the index $k$ refers to the number of creation operators equal to $c^{\dag}$ and $l$  to the number of annihilation operators equal to $c$ 
\footnote{The combinatorial factors result from the effect of combining of multiple elements of $\rho^{(n)}$ into one in the symmetrized form $\tilde{\rho}^{(n)}$, together with the change in normalization of the associated basis vectors.}.
With this definition, the symmetrized $n$-RDM $\tilde{\rho}^{(n)}$ has the same nonzero eigenvalues as $\rho^{(n)}$ and can therefore be used in place of $\rho^{(n)}$ for the calculation of effective cat sizes.

This projection onto the symmetric subspace results in a significant reduction in dimensionality, permitting calculations to be made for values of $n$ up to several hundred.  Matrix elements of $\tilde{\rho}_A^{(n)} - \tilde{\rho}_B^{(n)}$ are readily calculated for general forms of the amplitude spreading function $f(\theta)$ (see Appendix~\ref{app:RDMDerive}). A key component of these matrix elements are inner products between the states $\ket{\phi_{A,B}^N(\theta)}$ at different values of $\theta$, which yield factors of $\cos^N(\theta-\theta')$ and $\sin^N(\theta+\theta')$.
For large values of $N$ these functions can be approximated by delta functions.
 This simplifies the resulting integrals but removes any explicit $N$-dependence from the result (see Appendix~\ref{app:RDMDerive}).  
The matrix $\rho_A^{(n)} - \rho_B^{(n)}$ is then diagonalized and Eq.~(\ref{eq:PE_RDM}) evaluated to obtain the maximal probability of successfully distinguishing $\ket{\Psi^{(N)}_A}$ and $\ket{\Psi^{(N)}_B}$ with an $n$-particle measurement.
The effective cat size $C_{\delta}$ is then obtained by determining the minimum value of $n$ such that $P > 1 - \delta$, according to Eq.~(\ref{eq:catsize}).
When using the delta function approximation 
 for large $N$, since the total number of particles is unspecified, we evaluate the  \textit{relative} cat size, $C_\delta/N = 1/n_{\text{min}}$.

\begin{figure*}
\begin{center}
\begin{tabular}{p{0.5\textwidth}p{0.5\textwidth}}
\includegraphics[width=0.5\textwidth]{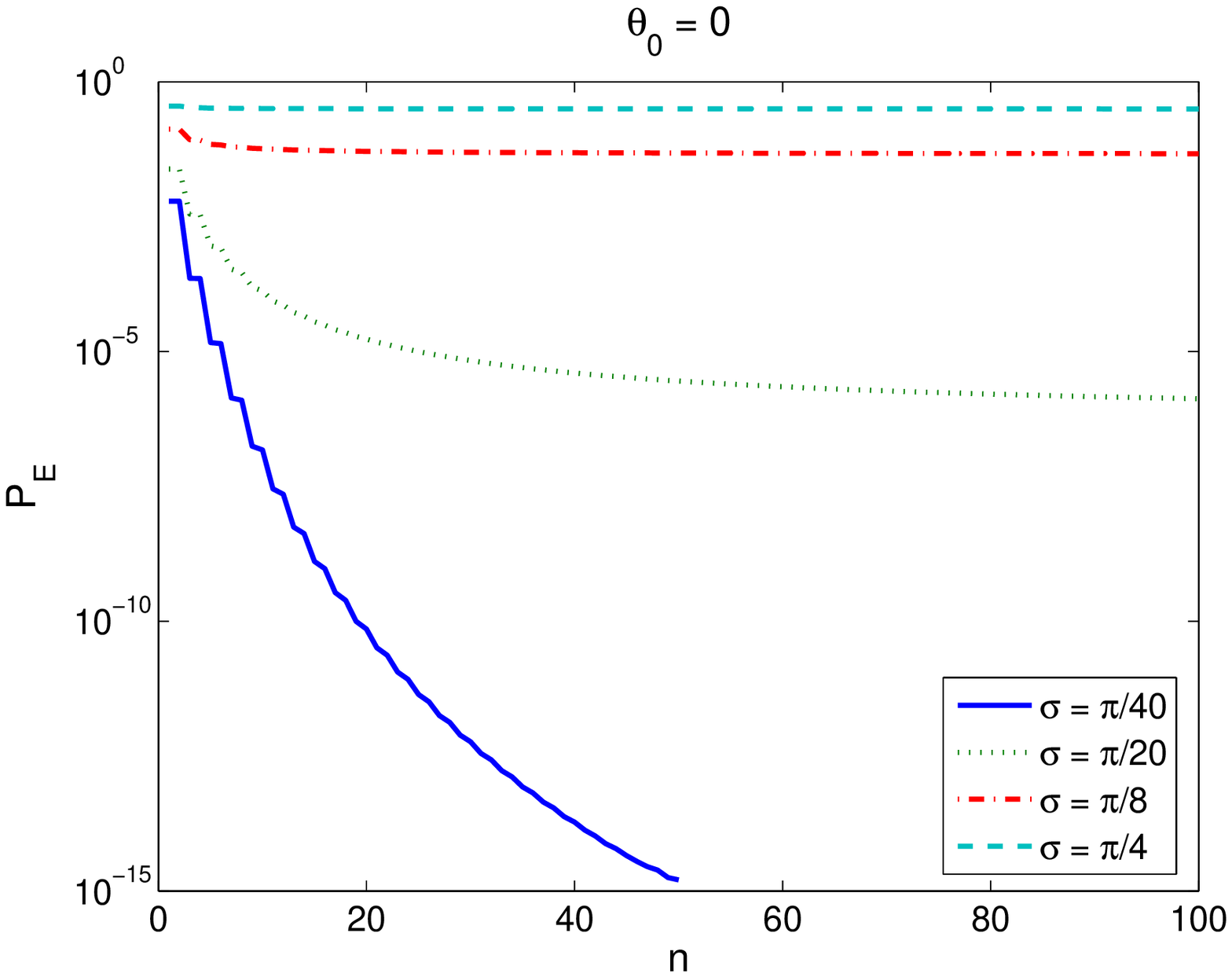}&
\includegraphics[width=0.5\textwidth]{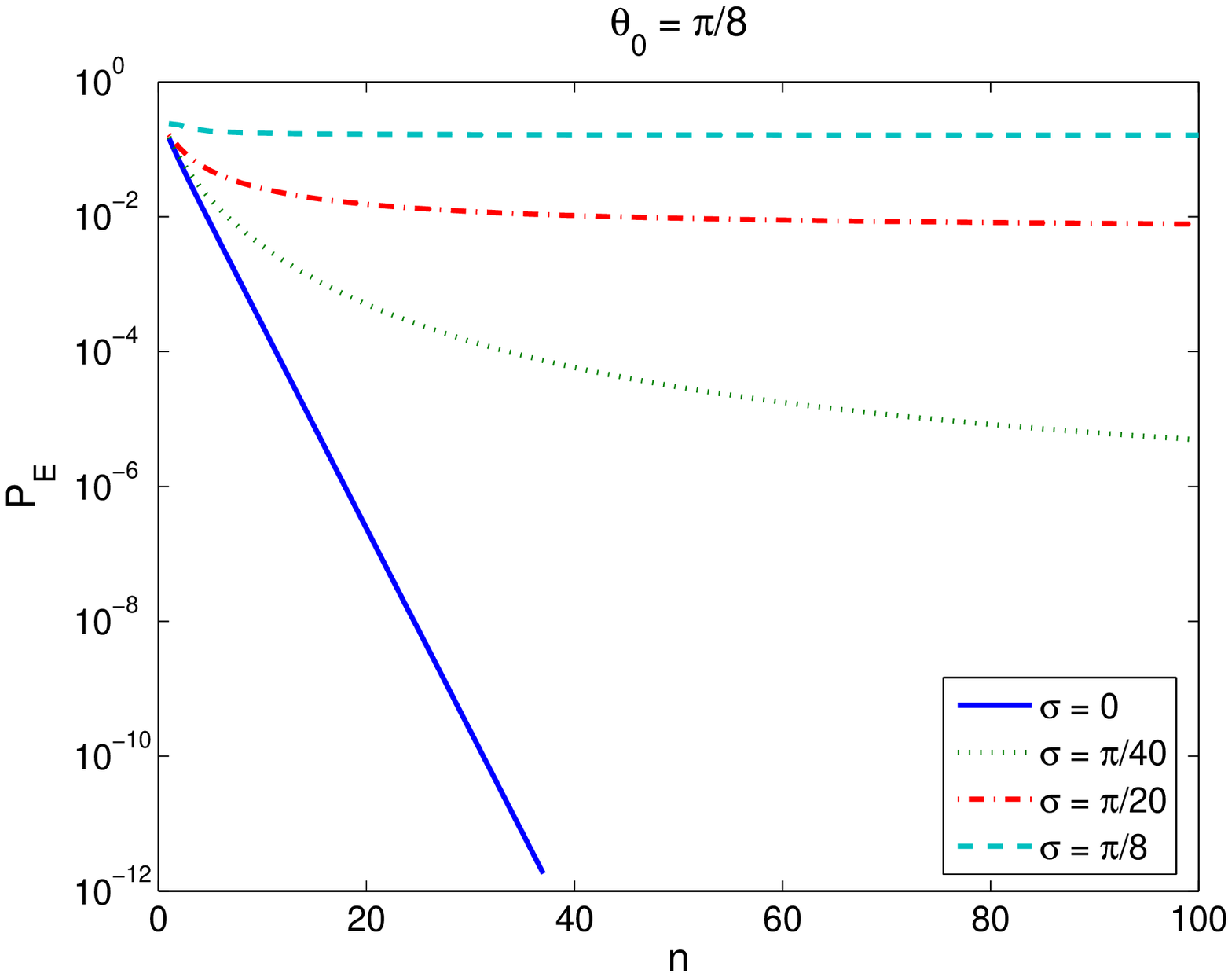}\\
\includegraphics[width=0.5\textwidth]{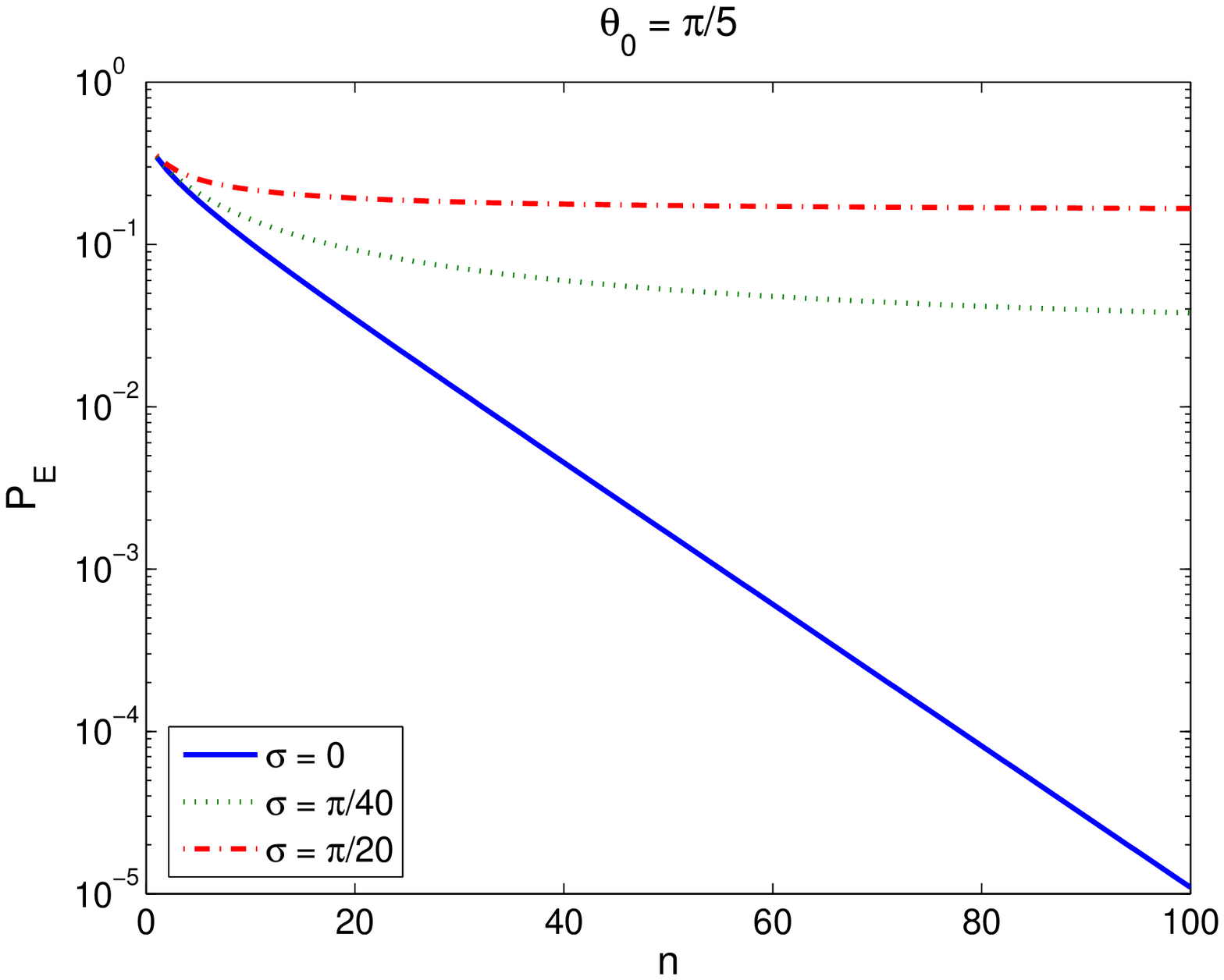}&
\includegraphics[width=0.5\textwidth]{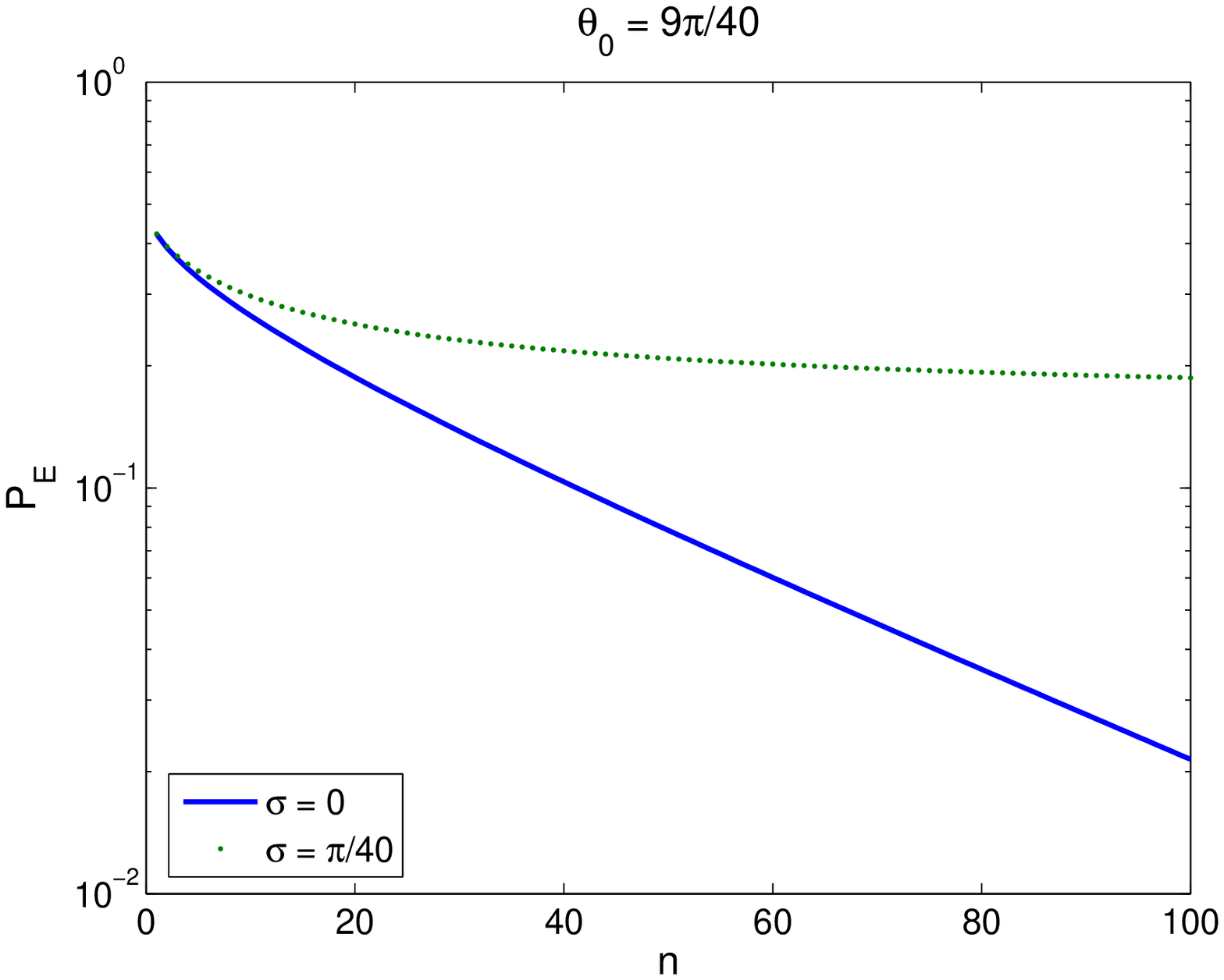}
\end{tabular}
\caption{(Color online.) Error probability, $P_E \equiv 1-P$, for distinguishing the two branches of the generalized cat state superposition Eq.~(\ref{eq:BECIntegratedCat}) when characterized by a gaussian amplitude spreading function $f(\theta)$, for various values of the gaussian parameters $\theta_0$ and $\sigma$.}
\label{fig:PEvsn}
\end{center}
\end{figure*}

\begin{figure*}
\begin{center}
\begin{tabular}{p{0.5\textwidth}p{0.5\textwidth}}
\includegraphics[width=0.5\textwidth]{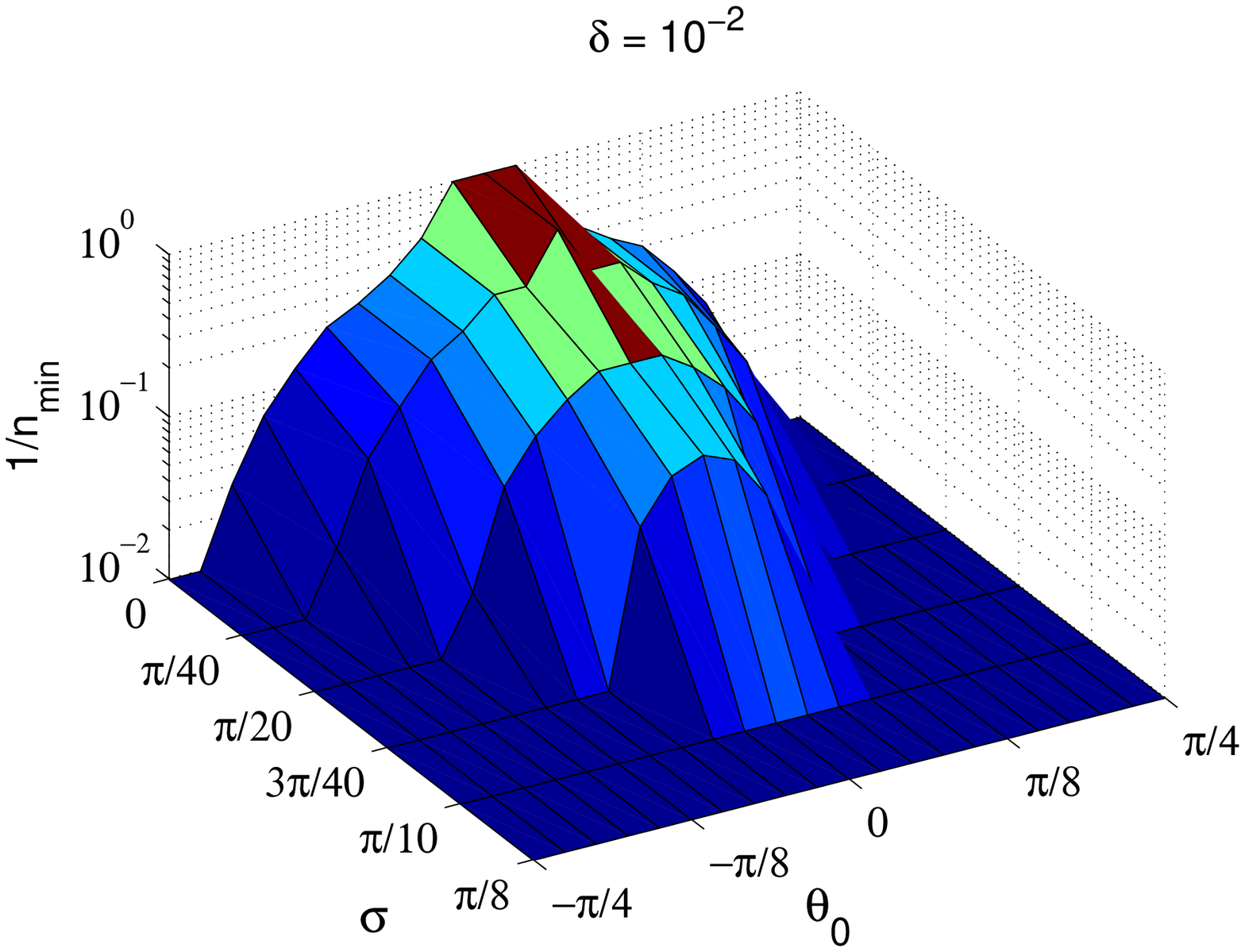}&
\includegraphics[width=0.5\textwidth]{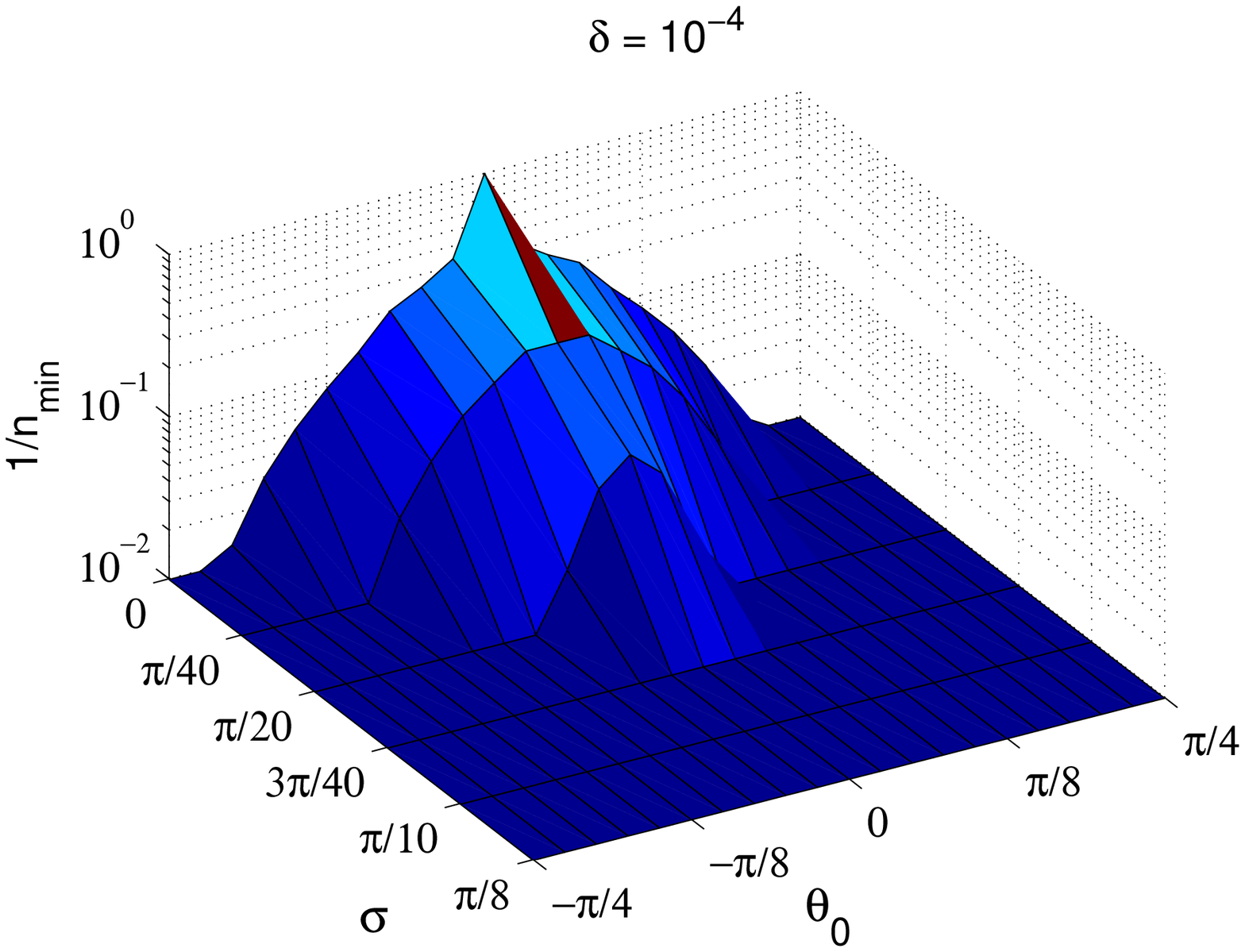}\\
\includegraphics[width=0.5\textwidth]{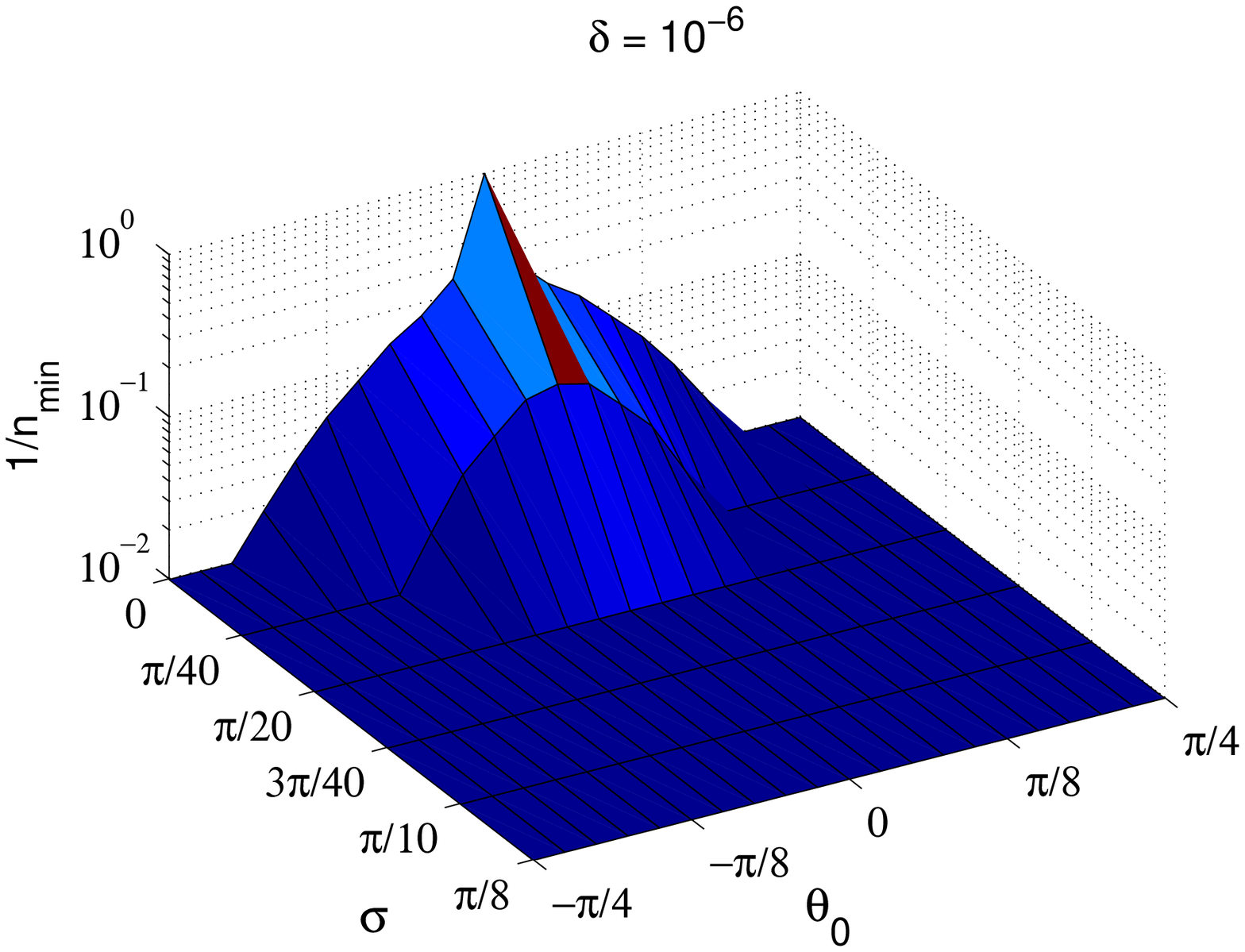}&
\includegraphics[width=0.5\textwidth]{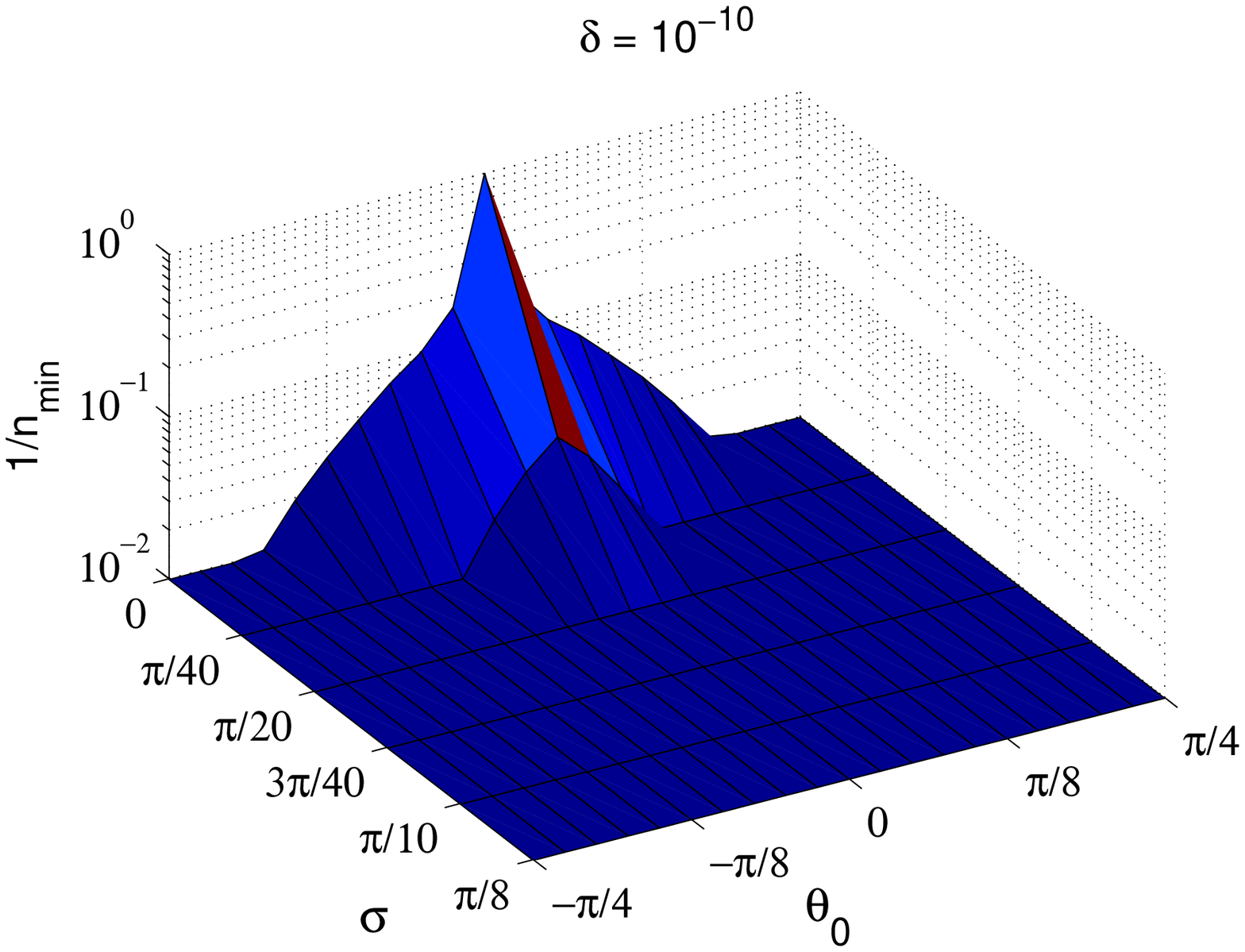}
\end{tabular}
\caption{(Color online.) Relative effective cat size $C_{\delta}/N\equiv 1/n_{min}$ as a function of the gaussian parameters $\theta_0$ and $\sigma$, for several values of desired precision $\delta$. All plots have a resolution of $\pi/40$ in both $\theta_0$ and $\sigma$.  Numerical calculations were made for $n \leq 100$, imposing a numerical cutoff of 0.01 on the value of $1/n_{min}$.}
\label{fig:RelativeCatSize}
\end{center}
\end{figure*}

Figures~\ref{fig:PEvsn} and \ref{fig:RelativeCatSize} show the results of calculations for a gaussian amplitude spreading function 
\begin{equation}
f(\theta) = \left ( 2\pi \sigma^2 \right )^{-1/4} e^{-\frac{\left(\theta - \theta_0 \right )^2}{4\sigma^2}}.
\label{eq:GaussianAmplitudeFunction}
\end{equation}
\noindent This form is convenient for a systematic analysis of the behavior of effective cat size with spread and overlap of the two branches since all matrix elements are analytic (see appendix~\ref{app:RDMDerive}).
The range of $\theta_0$ should be from $-\pi/2$ to $+\pi/2$ in order to encompass all relative phases and degrees of overlap/orthogonality.
Superposition states characterized by $\sigma = 0$ possess zero spread and reduce to the non-ideal states studied earlier in Ref.~\cite{DurSimonCirac2002}  that are characterized by the extent of non-orthogonality for  $\theta_0 > 0$.  
Figure~\ref{fig:PEvsn} shows the error probability $P_E=1-P$, plotted on a logarithmic scale as a function of $n$, for various values of the spread function parameters $\theta_0$ and $\sigma$.  We show $P_E$ rather than $P$, since the former allows a clearer analysis of the differences between results for $\sigma = 0$ and for $\sigma \ne 0$. The relative effective cat size $C_{\delta}/N\equiv1/n_{min}$ resulting from these probabilities is plotted as a function of $\theta_0$ and $\sigma$ for several different values of the precision parameter $\delta$ in Figure~\ref{fig:RelativeCatSize}.

Figure~\ref{fig:PEvsn} shows that while for all values of the parameters $\theta_0$ and $\sigma$ there is a generic increase in the probability $P$ for distinguishing the two branches of the cat state as $n$ increases (i.e., a \textit{decrease} in the error probability $P_E$), the nature of this decrease is strongly dependent on the actual values of $\theta_0$ and $\sigma$.  
For $\sigma=0$, the error is due entirely to non-orthogonality, as discussed in Ref.~\cite{DurSimonCirac2002} and Section~\ref{sec:catsize}.  Here, when $\theta_0=0$
the generalized superposition reduces to the ideal GHZ state and the error probability is zero, independent of $n$ (not shown in the bottom right panel since the logarithmic scale cannot accommodate $P_E = 0$).  When $\theta \neq 0$, the non-orthogonality makes the success probability increase more slowly with $n$, and hence the effective cat sizes in Figure~\ref{fig:RelativeCatSize} become smaller as $\theta_0$ approaches the value $\pm \pi/4$ at which the two branches $\ket{\Psi^{(N)}_A}$ and $\ket{\Psi^{(N)}_B}$ overlap completely.  In particular, for  strong overlap, $|\braket{\phi_A}{\phi_B}|^2 = 1 - \epsilon^2$ with $\epsilon \ll 1$ (outer limits of $\theta_0$ on $\sigma=0$ axis), we verify that the relative cat sizes are in accordance with the asymptotic scaling $\sim \epsilon^2$ established in Section~\ref{sec:catsize}.  This effect of non-orthogonality also acts when $\sigma > 0$, with the relative cat sizes also dropping off away from $\theta_0=0$. However now there is an additional decrease, due to the branches of the cat state getting ``smeared out'' and overlapping more as the width parameter $\sigma$ increases.  For all $\sigma$, we see that 
the effective cat size is largest for $\theta_0 = 0$, where the two branches $\ket{\phi_A^{(N)}(\theta_0)}$ and $\ket{\phi_B^{(N)}(\theta_0)}$ are orthogonal.

Detailed analysis of the dependence of the error probability $P_E$ on the width parameter $\sigma$  provides additional information.  When $\sigma=0$
and $\theta_0 \neq 0$,  consistent with the scaling shown in Section~\ref{sec:catsize} the error probability decreases exponentially with $n$ and asymptotically approaches zero as more particles are measured (solid blue lines in top right and bottom panels).  However, for $\sigma > 0$, we see that the decrease in the error probability is slower than exponential.  In fact it appears to never approach zero but is instead bounded below by some finite value, implying that the success probability is bounded away from unity. This derives from an important 
feature of this gaussian amplitude function $f(\theta)$ that is illustrated by comparing the overlap between $\ket{\Psi^{(N)}_A}$ and $\ket{\Psi^{(N)}_B}$ for different values of $\sigma$ and $\theta_0$. For example, at $\sigma=0$, $\theta \ne \pi/4$, the inner product between $\ket{\Psi^{(N)}_A}$ and $\ket{\Psi^{(N)}_B}$ goes to zero as $N \rightarrow \infty$, so that the two branches become orthogonal in the limit of an infinite number of particles, and one can therefore always tell them apart with arbitrarily high certainty by measuring enough of the particles 
(solid blue line). However, for  $\sigma > 0$, the overlap approaches a finite value as $N\rightarrow \infty$. In this situation it is not always possible to distinguish the two branches within a given precision, regardless of how many particles are measured -- even for $n=N$. This implies that $n_{\text{min}}$ is undefined for these extreme cases.
As noted in Section~\ref{sec:catsize}, we formally define $C_{\delta} = 0$ in these situations, with the additional understanding that $\ket{\Psi}$ is not really a meaningful cat state at all here.

This behavior for $\sigma > 0$ is consistent with the fact that the two branches $\ket{\phi^{(N)}_A(\theta)}$ and $\ket{\phi^{(N)}_B(\theta)}$ can be interchanged, either by transforming $\theta \rightarrow \pi/2 - \theta$, for $0\le\theta\le \pi/2$, or by first transforming $\theta \rightarrow -\pi/2-\theta$ and then changing sign, for $0\ge\theta\ge -\pi/2$.  Thus when the amplitude spread function $f(\theta)$ has support both inside and outside the region $-\pi/4 \le \theta \le +\pi/4$, some of  $\ket{\Psi}$ contributes to both branches $\ket{\Psi^{(N)}_A}$ and $\ket{\Psi^{(N)}_B}$, and the state cannot  be split into two disjoint branches.  Using Eqs.~(\ref{eq:phiAphiA})--(\ref{eq:phiAphiB}), it is also easy to see that for $\sigma = 0$, $\braket{\Psi_A^{(N)}}{\Psi_B^{(N)}} \rightarrow 0$ when $N \rightarrow \infty$, so that the branches become orthogonal and distinguishable in the thermodynamic limit, whereas for $\sigma \ne 0$, $\braket{\Psi_A^{(N)}}{\Psi_B^{(N)}}$ approaches a finite minimum value. This is the physical reason why two strongly overlapping branches cannot be distinguished to arbitrary high precision ($\delta \rightarrow 0$), even in the limit $N,n \rightarrow \infty$.  Detailed analysis of the support of the amplitude spread function will thus be very important for realistic estimates of cat size in physical systems involving superpositions of non-orthogonal states. 
 
This difference in behavior of success probability scaling for $\sigma = 0$ and for $\sigma > 0$ has a large effect on  the effective cat size.  Figure~\ref{fig:RelativeCatSize} shows the effective relative cat size $C_\delta/N$ for four different precision values, $\delta = 10^{-2}, 10^{-4}, 10^{-6}$ and $10^{-10}$.  It is evident that if $\delta$ is sufficiently small, the effective cat size does not depend too heavily on the exact value of $\delta$ when $\sigma = 0$. This is to be expected, since $1-P$ decreases exponentially with $n$ when $\sigma = 0$, and hence $n_{\text{\text{min}}}$ will only be proportional to $\log (1-P)$. However, when $\sigma > 0$, we see that the cat size can be significantly reduced or even vanish for a given system as we decrease the desired precision $\delta$. This illustrates the point made above, namely that states with $\sigma > 0$ become increasingly poor cat states as $\sigma$ increases and eventually are not cat states at all. It also provides a dramatic illustration of the general fact that the degree to which a superposition state can be viewed as a cat state is inherently dependent on the precision to which the implied measurements are made.

\subsection{Estimate of effective cat sizes from single-particle measurements}
\label{subsec:bosonic_mx1}

In all of the analysis so far, we have assumed that \textit{any} $n$-particle measurements can be made to distinguish the branches $\ket{A}$ and $\ket{B}$ of a cat state, including collective measurements in entangled bases. In practice, this is usually not feasible for large values of $n$.  From a practical perspective, it would therefore be desirable to have a definition of cat size which relies not on general $n$-particle measurements, but instead only makes use of measurements that can be put together from $n$ separate 1-particle measurements.

Allowing only those $n$-particle measurements that can be realized as a sequence of 1-particle measurements means that we restrict the corresponding POVM elements to be of the form
\begin{equation}
\mathcal{E} = \sum_{\{i\}} p_{\{i\}} E_{i_1}^{(1)} E_{i_2}^{(2)} \cdots E_{i_n}^{(n)},
\label{eq:SeparablePOVM}
\end{equation}
where each $E_{i_k}^{(k)} \equiv A_{i_k}^{(k)\dag} A_{i_k}^{(k)}$ acts on a single particle $k$ only, and where $p_i$ are positive numbers subject to the constraint that $\trace \mathcal{E} \le 1$. (Note that, unlike the situation in Sections~\ref{sec:catsize} and \ref{subsec:bosonic_effectivecatsizes}, the POVM elements here act each on only a single particle, and the superscript index $(k)$ in parentheses therefore labels the particle that each operator acts on, not the \textit{number} of particles it acts on.) This means that the POVM elements must be \textit{separable}. Furthermore, to ensure that the measurements can be realized as a \textit{sequence} of 1-particle measurements, it must be possible to write express the POVM elements in such a way that $E^{(k)}_{i_k}$ only depends on $E^{(l)}_{i_l}$ for $l < k$ but not for $l > k$. To find the maximum probability $P$ of successfully distinguishing the branches $\ket{A}$ and $\ket{B}$ of a cat state using such measurements, we would then need to maximize Eq.~(\ref{eq:SuccessProb}) with $E_A^{(n)}$ and $E_B^{(n)}$ subject to the above constraints. Unfortunately, we know of no efficient way to do this. In particular, deciding whether a given POVM is separable as in Eq.~(\ref{eq:SeparablePOVM}) is known to be an $NP$-hard problem \cite{Gurvits2003}.

However, if we restrict ourselves to a very simple case, namely to superposition states where each of the two branches of the cat state are themselves product states, not only is the optimal measurement strategy using a sequence of $n$ one-particle measurements known, but it even performs equally well as the optimal general $n$-particle measurement. To show this we adapt the techniques used in~\cite{AcinBaganBaig2005}. In that work, one is given $n$ copies of a quantum system, all prepared in one of two states $\ket{\psi_A}$ and $\ket{\psi_B}$ and asked to tell which one (note that \cite{AcinBaganBaig2005} uses $0,1$ rather than $A,B$). The \textit{joint} state of all $n$ copies is then either $\ket{\psi_A}^{\otimes n}$ or $\ket{\psi_B}^{\otimes n}$, and the corresponding density matrix is $\rho_{\gamma}^{\otimes n} \equiv (\ket{\psi_{\gamma}}\bra{\psi_{\gamma}})^{\otimes n}$, for $\gamma =A,B$. One assumes prior probabilities $q_A$ and $q_B = 1-q_A$ that the correct state is $\ket{\psi_A}$ and $\ket{\psi_B}$, respectively. The maximum possible probability of guessing the right state would in general consist of making an optimally chosen collective $n$-party measurement (i.e., possibly in an entangled basis) on the $n$ copies. However, it is shown that by measuring only a single copy at a time and choosing each measurement according to a protocol that effectively amounts to Bayesian updating of the priors $q_A$ and $q_B$ based on the outcome of the previous measurement, one can obtain a success probability which is equal to the maximum one for a general $n$-party measurement.

In our case, we are trying to ascertain whether a \textit{single} system consisting of $N$ subsystems is in a state $\ket{\Psi_A}$ or another state $\ket{\Psi_B}$, where these states are known to be product states with respect to the $N$ subsystems. We can therefore write
\begin{equation}
\ket{\Psi_{\gamma}} \equiv \ket{\psi^{(1)}_{\gamma}} \otimes \ket{\psi^{(2)}_{\gamma}} \otimes \cdots \otimes \ket{\psi^{(N)}_{\gamma}}
\label{eq:Psigamma}
\end{equation}
where $\gamma = A,B$ and $\ket{\psi^{(k)}_{\gamma}}$ is the state of particle number~$k$, and we assume that we will measure the first $n$ particles. This is equivalent to a generalization of~\cite{AcinBaganBaig2005} to a situation where not all the copies of the system under study are the same, but where each ``copy''~$k$ is in one of two states $\ket{\psi^{(k)}_{\gamma}}$ for $\gamma=A$ or $B$, and where $\gamma$ is the same for each $k$, and the task is to determine the value of $\gamma$, by only measuring $n$ of the ``copies''. We will now show that the conclusion of~\cite{AcinBaganBaig2005} still holds in this case, namely that the performing a sequence of $n$ optimal one-particle measurements with Bayesian updating between each measurement gives the same probability of success as the best collective $n$-particle measurement. We will use a slightly different approach than~\cite{AcinBaganBaig2005}, using 1-particle reduced density matrices instead of single-particle state vectors, since this approach is more readily generalizable to indistinguishable particles.

Following the notation of \cite{AcinBaganBaig2005}, we will here write the states $\ket{\psi^{(k)}_A}$ and $\ket{\psi^{(k)}_B}$ of particle $k$ in the branches $\ket{\Psi_A}$ and $\ket{\Psi_B}$ respectively as
\begin{equation}
\ket{\psi^{(k)}_{\gamma}} \equiv \cos \theta_k \ket{x_k} + \left (-1 \right )^a \sin \theta_k \ket{y_k}
\label{eq:Psikgamma}
\end{equation}
where $a=0$ for $\gamma=A$ and $a=1$ for $\gamma=B$, and $\ket{x}$ and $\ket{y}$ are two basis vectors in the state space of particle $k$ chosen such that this relation is valid (this is always possible). The corresponding reduced density matrix with respect to particle $k$ in the $\{\ket{x}$,$\ket{y}\}$ basis are then
\begin{equation}
\begin{split}
\rho^{(k)}_{\gamma} &=
\begin{pmatrix}
\cos^2 \theta_k & (-1)^a \cos \theta_k \, \sin \theta_k \\
(-1)^a \cos \theta_k \, \sin \theta_k & \sin^2 \theta_k
\end{pmatrix} \\
&=
\begin{pmatrix}
\cos^2 \theta_k & \frac{(-1)^a}{2} \sin 2\theta_k \\
\frac{(-1)^a}{2} \sin 2\theta_k & \sin^2 \theta_k
\end{pmatrix}
\end{split}
\label{eq:rhokgamma}
\end{equation}
(note here that the superscript $k$ again refers to the particle to which the RDM belongs, not the number of particles described by the RDM, which in this case is just $1$.)
If we now let the probability, prior to measuring particle $k$, of the state being $\ket{\psi^{(k)}_{\gamma}}$ be $q^{(k)}_{\gamma}$, then the measurement which produces the highest probability of successfully identifying the correct state, is a projective measurement in the basis in which the matrix $\Gamma^{(k)} \equiv q^{(k)}_0 \rho^{(k)}_0 - q^{(k)}_1 \rho^{(k)}_1$ is diagonal (\cite{Helstrom1976, Helstrom1969}). The conclusion $\gamma=A$ is associated with the eigenspaces with positive eigenvalues of $\Gamma^{(k)}$, while $\gamma=B$ corresponds to the eigenspaces with negative eigenvalues. In the basis $\{\ket{x_k}$,$\ket{y_k}\}$, the matrix $\Gamma^{(k)}$ is:
\begin{equation}
\Gamma^{(k)} =
\begin{pmatrix}
(q^{(k)}_A-q^{(k)}_B)\cos^2\theta & \frac{1}{2} (q^{(k)}_A + q^{(k)}_1) \sin 2\theta \\
(q^{(k)}_A + q^{(k)}_B)\sin 2 \theta & (q^{(k)}_A-q^{(k)}_B)\sin^2\theta
\end{pmatrix}
\label{eq:GammakNonDiagonalBasis}
\end{equation}
and is diagonalized by
\begin{equation}
\mathbf{U}(\phi_k) = 
\begin{pmatrix}
\cos \phi_k & \sin \phi_k \\
-\sin \phi_k & \cos \phi_k
\end{pmatrix}
\end{equation}
with
\begin{align}
\sin 2\phi_k &= \frac{q^{(k)}_A + q^{(k)}_B}{R_k} \, \sin 2\theta_k = \frac{1}{R_k} \, \sin 2\theta_k\\
\cos 2\phi_k &= \frac{q^{(k)}_A - q^{(k)}_B}{R_k} \, \cos 2\theta_k \\
R_k &= \sqrt{(q^{(k)}_A + q^{(k)}_B)^2 - 4q^{(k)}_A q^{(k)}_B \cos^2 2\theta_k} \nonumber \\
&= \sqrt{1 - 4q^{(k)}_A q^{(k)}_B \cos^2 2\theta_k},
\label{eq:Rk}
\end{align}
resulting in eigenvalues
\begin{equation}
\lambda^{(k)}_{A,B} \equiv \frac{1}{2} \left ( q^{(k)}_A - q^{(k)}_B \right ) \pm \frac{1}{2}R_k\, .
\end{equation}

The outcome $E^{(k)}_A$ is associated with the eigenspace of $\Gamma^{(k)}$ corresponding to the eigenvalue $\lambda^{(k)}_A$, which is the first eigenvector in the diagonal basis. In the basis used in Eq.~(\ref{eq:GammakNonDiagonalBasis}), we then have
\begin{equation}
\begin{split}
E^{(k)}_A &= U(\phi_k)^{\dag}
\begin{pmatrix}
1 & 0 \\
0 & 0
\end{pmatrix}
U(\phi_k) \\
&=
\begin{pmatrix}
\cos^2 \phi_k & \frac{1}{2} \sin 2\phi_k \\
\frac{1}{2} \sin 2\phi_k & \sin^2 \phi_k
\end{pmatrix}
\end{split}
\end{equation}
Combining this with Eq.~(\ref{eq:rhokgamma}) gives us the conditional probabilities $P(E^{(k)}_A | \gamma) = \trace \left ( E^{(k)}_A \rho^{(k)}_{\gamma} \right )$ of obtaining the outcome $E^{(k)}_{A}$ when measuring particle $k$, \textit{given} that the initial state of the joint system was $\ket{\Psi_{\gamma}}$:
\begin{align}
P(E^{(k)}_A | A) &=
\frac{1}{2} + \frac{1}{2R_k} \left ( 1 - 2q^{(k)}_B \cos^2 2\theta_k \right )
\label{eq:PEAGivenA}\\
P(E^{(k)}_A | B)
&= \frac{1}{2} - \frac{1}{2R_k} \left ( 1 - 2q^{(k)}_A \cos^2 2\theta_k \right ) \, .
\label{eq:PEAGivenB}
\end{align}
The corresponding probabilities of obtaining $E^{(k)}_B = \openone - E^{(k)}_A$ are then
\begin{align}
P(E^{(k)}_B | A)
&= \frac{1}{2} - \frac{1}{2R_k} \left ( 1 - 2q^{(k)}_B \cos^2 2\theta_k \right )
\label{eq:PEBGivenA} \\
P(E^{(k)}_B | B)
&= \frac{1}{2} + \frac{1}{2R_k} \left ( 1 - 2q^{(k)}_A \cos^2 2\theta_k \right )
\label{eq:PEBGivenB}
\end{align}
Using Eqs.~(\ref{eq:PEAGivenA}) and (\ref{eq:PEBGivenB}), the probability of successfully identifying the state $\ket{\psi^{(k)}_{\gamma}}$ after measuring particle~$k$ (conditional upon ealier measurements yielding the priors $q^{(k)}_A$ and $q^{(k)}_B$) is
\begin{equation}
\begin{split}
P_k &\equiv q^{(k)}_A P(E^{(k)}_A | A) + q^{(k)}_B P(E^{(k)}_B | B) \\
&= \frac{1}{2} + \frac{1}{2} R_k
\end{split}
\label{eq:Pk}
\end{equation}

To find the overall success probability of the procedure, we need to evaluate what the posterior probabilities for $\gamma = A$ and $\gamma = B$ are after measuring each particle. These will then serve as the prior probabilities $q^{(k+1)}_A$ and $q^{(k+1)}_B$ for the next measurement, and the overall success probability will be the probability of obtaining the correct result at the very \textit{last} measurement. The outcome of this measurement will be used as the indicator of what the initial state was.
Similar to~\cite{AcinBaganBaig2005}, we show in Appendix~\ref{app:PosteriorProbs} that one of the posterior probabilities $q^{(k+1)}_{\gamma}$ will be equal to the success probability $P_k$ of the $k$'th measurement, while the other will be the error probability $\overline{P}_k = 1-P_k$.
We then know that either $q^{(k+1)}_A = P_k$ and $q^{(k+1)}_B = 1 - P_k$ if the outcome $E^{(k)}_A$ was obtained, or vice versa if the outcome $E^{(k)}_B$ was obtained. To simplify the notation in the following, we define $c_k^2 \equiv \cos^2 \theta_k = |\braket{\psi^{(k)}_A}{\psi^{(k)}_B}|^2$.
Combining Eqs.~(\ref{eq:Pk}) and (\ref{eq:Rk}) we can then establish the recursive relation
\begin{equation}
\begin{split}
R_k &= \sqrt{1 - 4 P_{k-1} (1-P_{k-1}) c_k^2}
\end{split}
\end{equation}
whose solution is
\begin{equation}
R_k = \sqrt{1 - 4q^{(1)}_A q^{(1)}_B \prod_{l=1}^k c_l^2}
\end{equation}
From this we see that the probability of obtaining the correct result when measuring particle number~$n$, the last of the $n$ particles to be measured, and hence the overall probability of success, is equal to
\begin{equation}
P_n = \frac{1}{2} + \frac{1}{2}\sqrt{1 - 4 q_A q_B \prod_{k=1}^n c_k^2}
\label{eq:1ParticleSuccessProbability}
\end{equation}
where $q_A \equiv q^{(1)}_A$ and $q_B \equiv q^{(1)}_B$ are the priors before the start of the whole measurement series. When we apply this to measuring cat size, we assume equal weight for the two branches, so that $q_A = q_B = 1/2$, and $P_n = 1/2 + 1/2\sqrt{1-\prod_k c_k^2}$. Now if we employ the same reasoning as went into deriving Eq.~(\ref{eq:SuccessProbability}) for the success probability of the optimal \textit{collective} $n$-particle measurement, we easily obtain that this is identical to $P_n$ in Eq.~(\ref{eq:1ParticleSuccessProbability}). Hence, when the branches are product states, a sequence of single-particle measurements with Bayesian updating has the same success probability as the optimal $n$-particle measurement.

The above discussion was carried out entirely in terms of distinguishable particles. The result generalizes partly to bosonic system, but not entirely. The result holds if each of the branches are single-mode Fock states with all $N$ particles in the same mode, i.e. $\ket{\Psi_A} = (a^{\dag})^N \ket{0}/N!$ and $\ket{\Psi_B} = (b^{\dag})^N \ket{0}/N!$, where the modes created by $a^{\dag}$ and $b^{\dag}$ are not necessarily orthogonal. If we then write $a^{\dag} = \cos \theta \, c_x^{\dag} + \sin \theta \, c_y^{\dag}$ and $b^{\dag} = \cos \theta \, c_x^{\dag} - \sin \theta \, c_y^{\dag}$ in analogy with Eq.~(\ref{eq:Psikgamma}), where $c_x^{\dag}$ and $c_y^{\dag}$ are creation operators for orthogonal modes $x$ and $y$, the bosonic $n$-RDMs that we obtain using the techniques from Section~\ref{sec:bosonic} are identical to those we obtain for distinguishable particles using Eqs.~(\ref{eq:Psigamma}) and (\ref{eq:Psikgamma}). Furthermore, the action of the optimal measurements obtained in the bosonic case can (at least in principle) be realized through Kraus operators consisting of a single annihilation operator for each measurement. This simply annihilates a single boson without changing the joint state of the system in any other way.
Hence all conclusions obtained for distinguishable particles carry over to the bosonic case in this situation.

However, if each branch is a more general Fock state with more than one occupied mode, i.e. of the form
\begin{align}
\ket{A} &\propto \prod_k a_k^{\dag} \, \ket{0} & \ket{B} &\propto \prod_k b_k^{\dag} \, \ket{0}
\label{eq:SingleParticleFockStateBranches}
\end{align}
where $a_k^{\dag}$ and $a_{k'}^{\dag}$ may create particles in different modes (not necessarily orthogonal) when $k\ne k'$, then the single-particle Bayesian updating measurement protocol derived above for distinguishable partcles cannot even be implemented. Since the particles are not distinguishable and cannot be addressed individually, there is no way to associate a single value of $k$ with each measurement, and hence no way to optimize each single-particle measurement in the way we did above. Furthermore, if the modes associated with different $a_k^{\dag}$ or $b_k^{\dag}$ are not orthogonal, then the branches in Eq.~(\ref{eq:SingleParticleFockStateBranches}) in fact contain entanglement between modes, and measuring one particle will therefore change the state of the remaining system and affect subsequent measurements. Hence the protocol described in this section only works for bosonic systems when each branch is a Fock state with all particles in a \textit{single} mode.

\section{$n$-RDM entropies and related cattiness measures}
\label{sec:entropies_disconnectivity}

We now analyze the von~Neumann entropy of the $n$-RDM $\rho^{(n)}$ and show that this provides insight into how meaningful it is to treat the state $\ket{\Psi}$ of Eq.~(\ref{eq:BECIntegratedCat}) as a two-branch cat state. Calculating the entropy of the $n$-RDM also allows us to compare our cat-size measure to an earlier one, the so-called ``disconnectivity'' introduced by Leggett \cite{Leggett1980}.

The von~Neumann entropy of a density matrix $\rho$ is given as
\begin{equation}
S = - \trace \left [ \rho \, \ln \rho \right ] \equiv -\sum_i \rho_i \, \ln \rho_i,
\end{equation}
where $\{ \rho_i \}$ are the eigenvalues of $\rho$.  Analogous to the Shannon entropy of a probability distribution, this quantity tells us how much information is encoded in the knowledge of the physical system represented by the density matrix. Equivalently, it can be viewed as the minimum amount of ignorance, we can have about the outcome of any measurement on a system represented by a given density matrix, where the minimization is over all possible measurements encompassed by the density matrix, {\it i.e.}, all possible $n$-particle measurements in the case of an $n$-RDM.
To evaluate the von~Neumann entropy $S_n$ characterizing $n$-particle measurements on a cat state we need the $n$-RDM $\tilde{\rho}^{(n)}$ of the full state $\ket{\Psi}$ and not just that of the individual branches. This is calculated for the states of Eq.~(\ref{eq:BECIntegratedCat}) in Appendix~\ref{app:RDMDerive}.

Before analyzing the entropy of $\tilde{\rho}^{(n)}$ for Eq.~(\ref{eq:BECIntegratedCat}), we first summarize how the entropy should scale for general classes of cat-like and non-cat-like states.
In general, for an experiment that has $d$ equally likely outcomes, the entropy of the probability distribution is simply $\ln d$. If not all outcomes are equally likely, then the entropy $S$ will be less than $\ln d$. Therefore, if the probability distribution of a measurement has entropy $S$, then the measurement must have at least $e^S$ distinct outcomes. This further means that, since the von~Neumann entropy of a density matrix is the \textit{minimum} entropy of any measurement describable by that density matrix, any measurement on a system whose von~Neumann entropy is $S$ must also have at least $e^S$ distinguishable outcomes.

For a perfect cat state, schematically of the form $\ket{\psi} = 1/\sqrt{2} \left ( \ket{0}^{\otimes N} + \ket{1}^{\otimes N} \right )$ with $\braket{0}{1} = 0$, this implies that the $n$-RDM of the system will have a 
von~Neumann entropy $S_n=\ln 2$, independent of $n$, until $n=N$ where $S_N=0$. If we make a single-particle measurent in the $\{ \ket{0}, \ket{1} \}$ basis, the outcomes $\ket{0}$ and $\ket{1}$ are both equally likely, so the entropy of that measurement is $\ln 2$. Unless we measure all $N$ particles however, measuring more particles gives us no additional information, since measuring just one particle completely collapses the system into one of its branches, and hence the entropy of the $n$-RDM for all $n < N$ is equal to $\ln 2$.
For a ``poor'' cat state, e.g., one of the form $\ket{\psi} \sim \ket{\phi_0}^{\otimes N} + \ket{\phi_1}^{\otimes N}$ with $\braket{\phi_0}{\phi_1} \ne 0$, we cannot distinguish the two branches perfectly with an $n$-particle measurement.  
One can show that the von~Neumann entropy in this case will be less than $\ln 2$. However, as we measure more and more particles, the branches become more and more distinguishable as they approach orthogonality in the thermodynamic limit.  Hence the von~Neumann entropy will asymptotically approach $\ln 2$ as $n$ grows.  It will then decrease to zero again, in a symmetric fashion, as $n$ approaches $N$, as more and more information about the coherence of the branches becomes available.

Unlike such cat-like states, the entropy of the $n$-RDM of completely generic (pure) states will usually not level out as $n$ increases. For a generic state, measuring $n$ particles is not likely to tell us very much about the effect of adding an $n+1$'th particle to the measurement. Therefore, the number of distinguishable outcomes will usually keep increasing with $n$, until it reaches $\sim N/2$. At that point, we will start gaining enough phase information that the entropy will start decreasing again. At this point, the number of particles that we are tracing out becomes smaller than the number of particles we are keeping, so the entropy can increase no further, and instead drops steadily, until it reaches zero at $n=N$ (in a pure state).

\begin{figure*}
\begin{center}
\begin{tabular}{p{0.5\textwidth}p{0.5\textwidth}}
\includegraphics[width=0.5\textwidth]{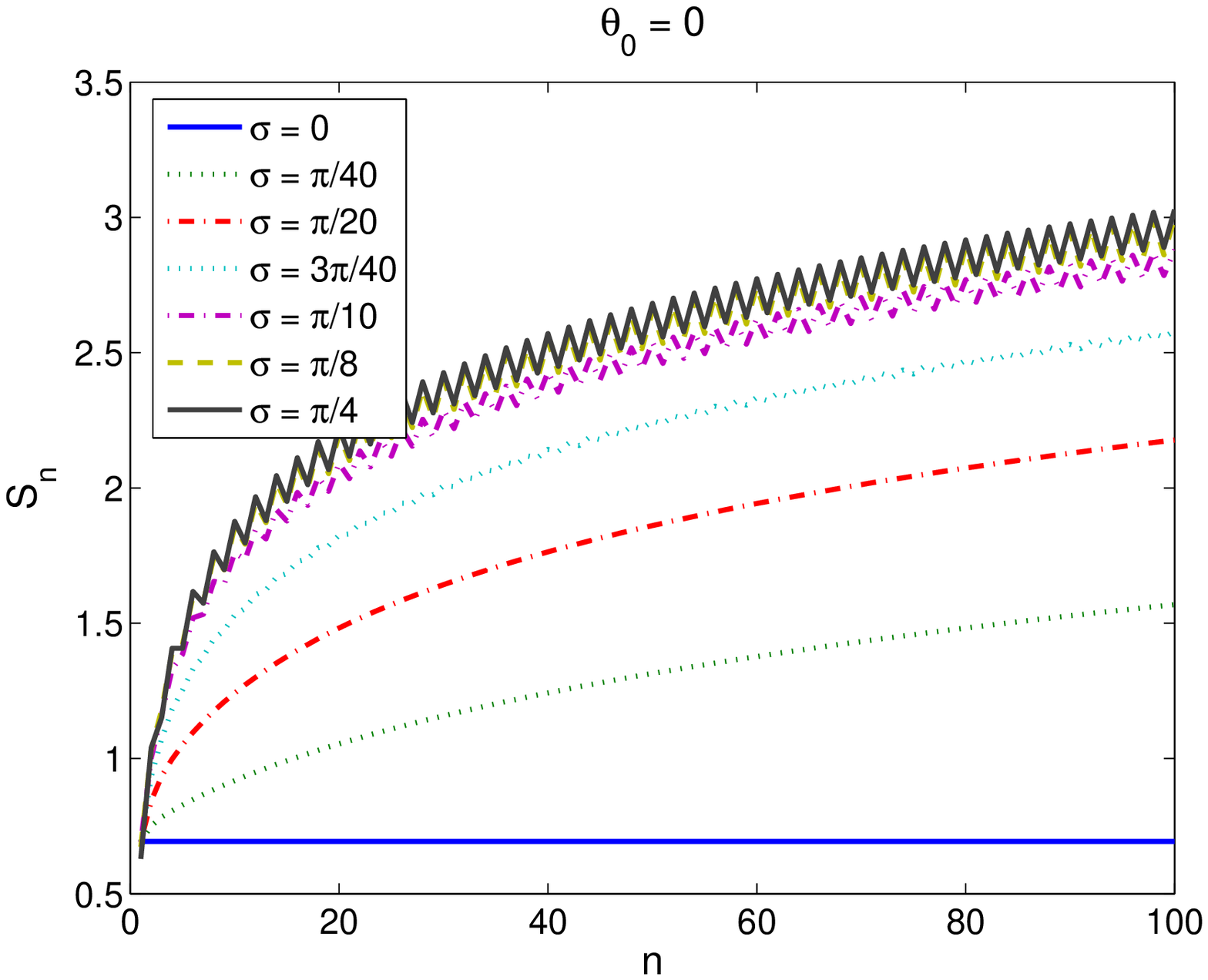}&
\includegraphics[width=0.5\textwidth]{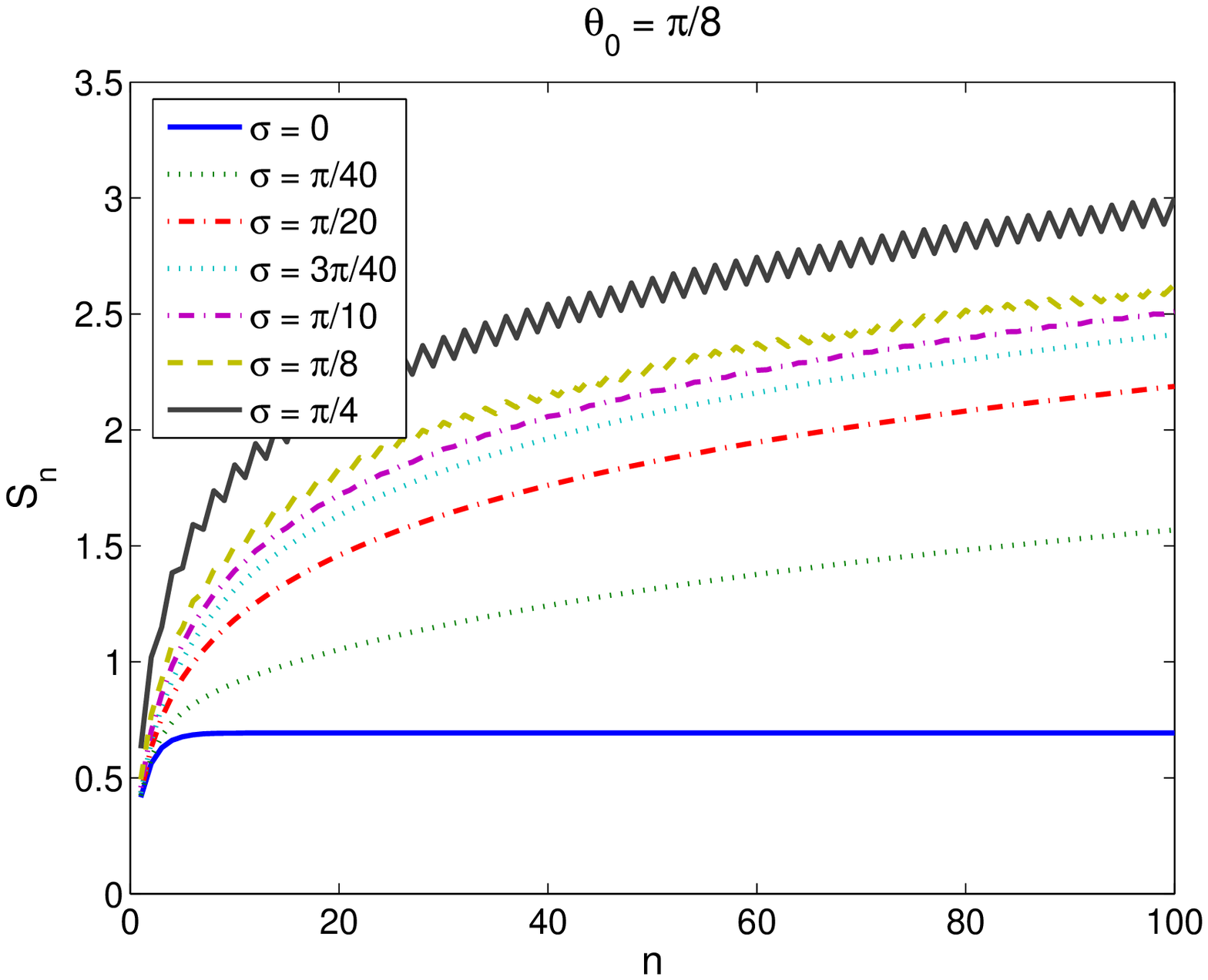}\\
\includegraphics[width=0.5\textwidth]{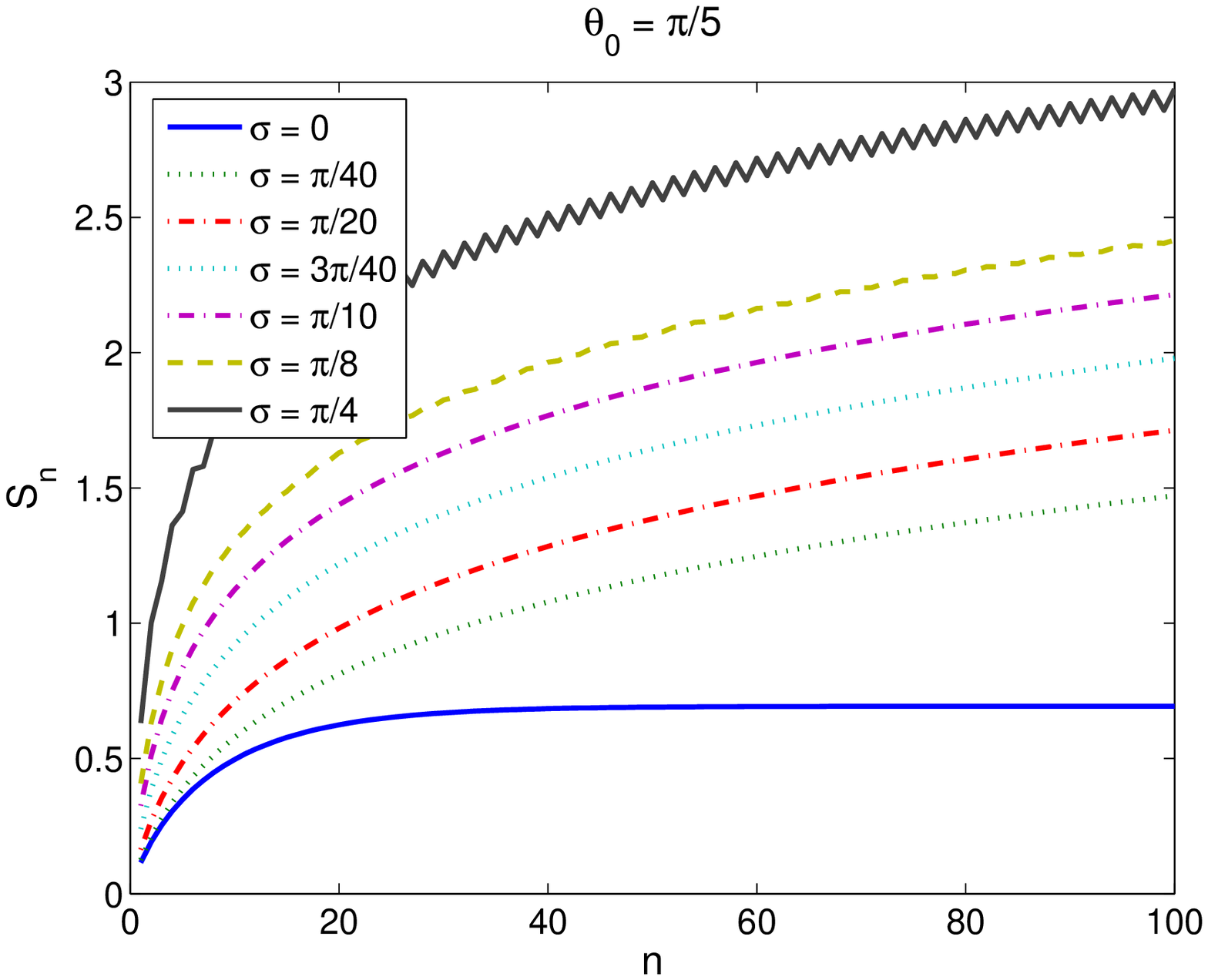}&
\includegraphics[width=0.5\textwidth]{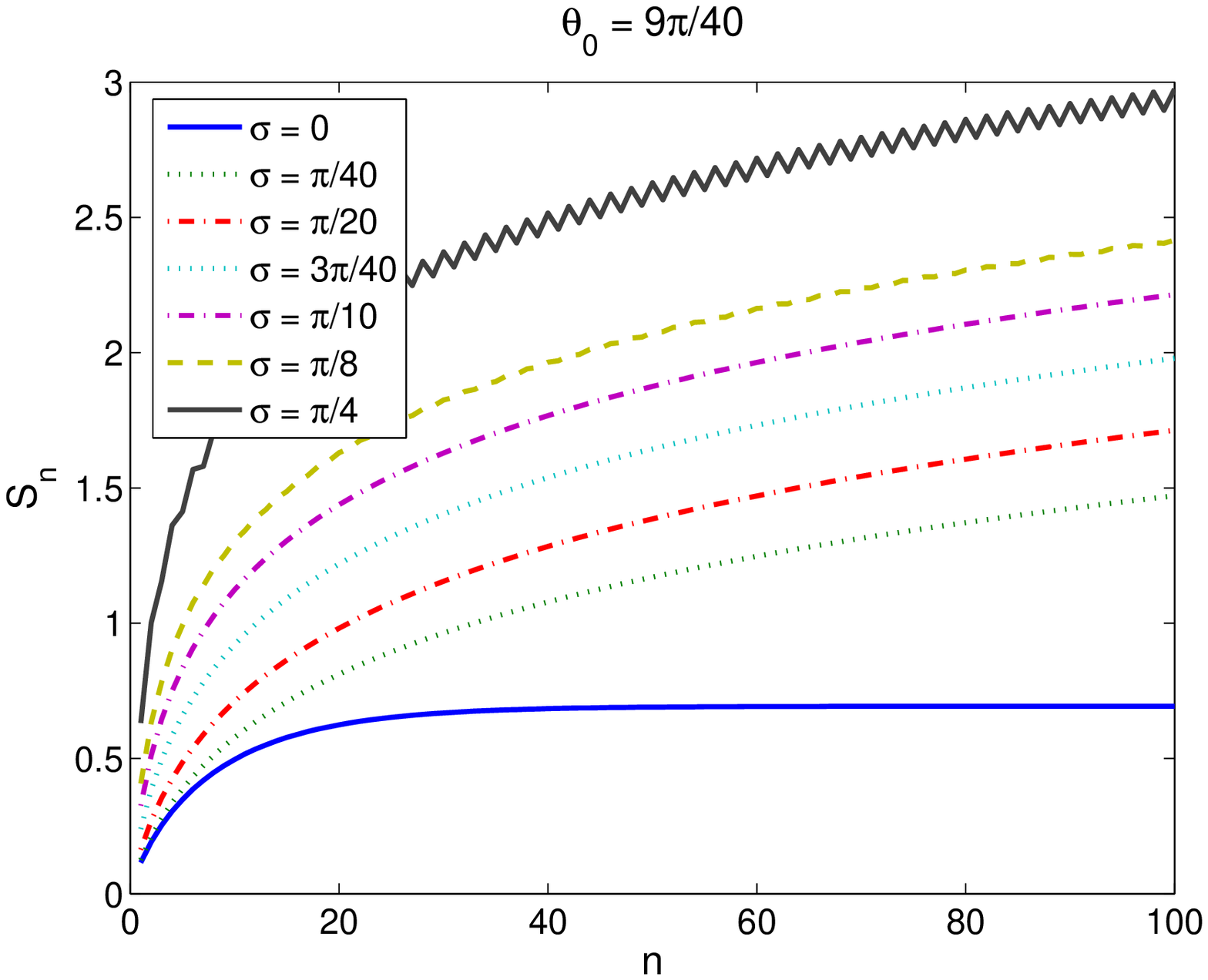}
\end{tabular}
\caption{(Color online.) von Neumann entropies of $n$-RDMs for various values of $\theta_0$ and $\sigma$, 
evaluated in all cases with $n \ll N$.}
\label{fig:nRDMEntropies}
\end{center}
\end{figure*}

\begin{figure*}
\begin{center}
\begin{tabular}{p{0.5\textwidth}p{0.5\textwidth}}
\includegraphics[width=0.5\textwidth]{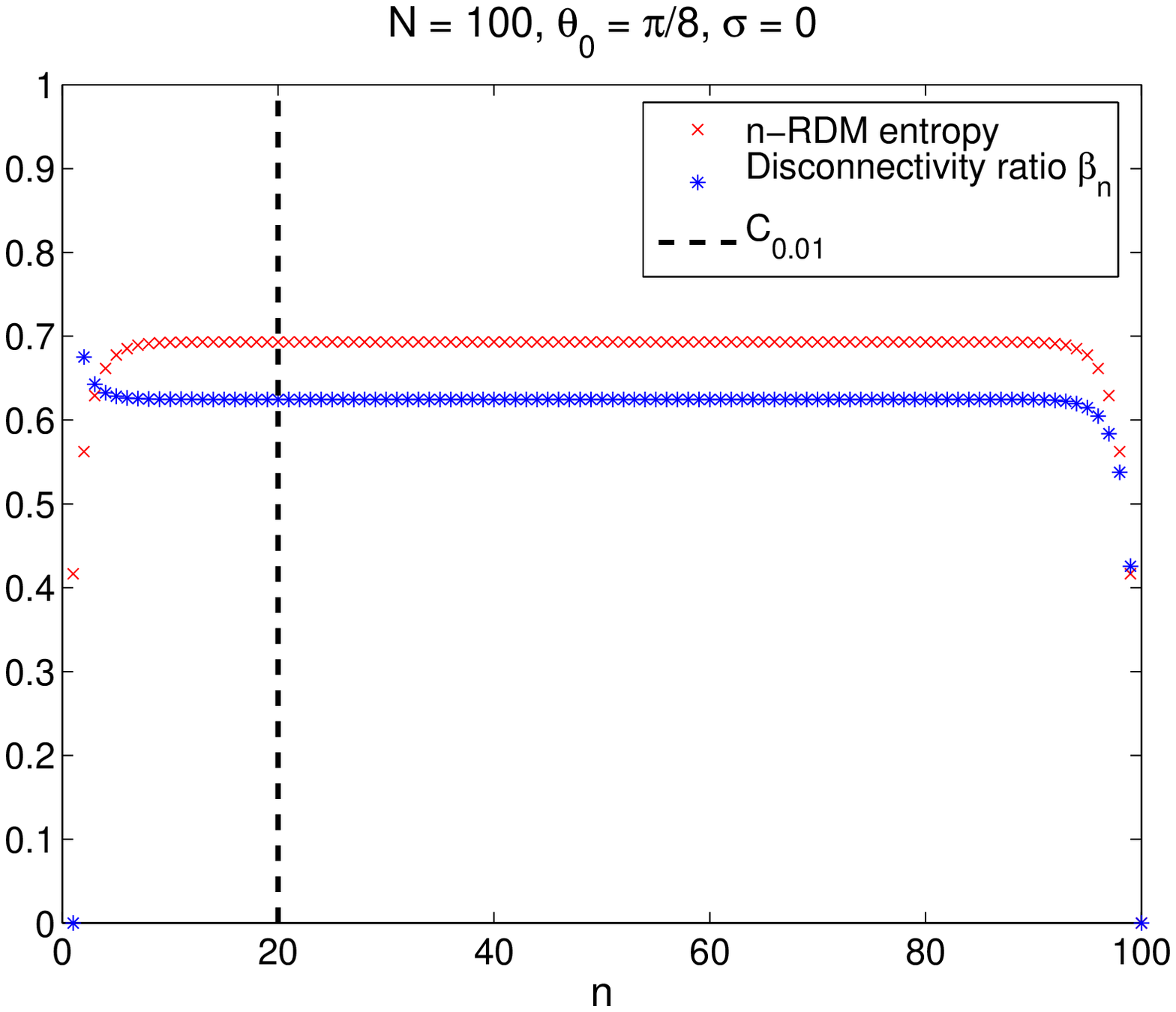} &
\includegraphics[width=0.5\textwidth]{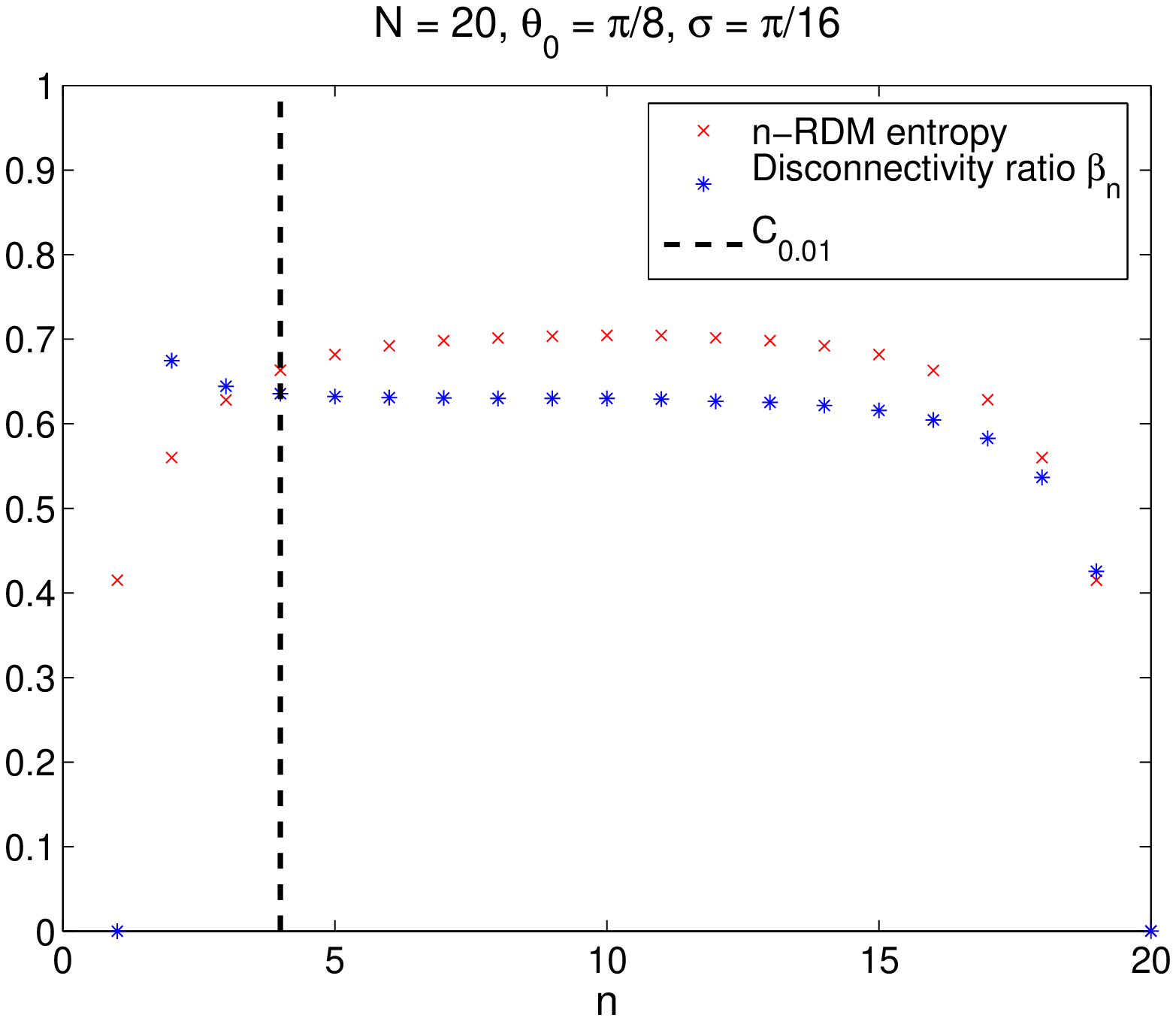} \\
\includegraphics[width=0.5\textwidth]{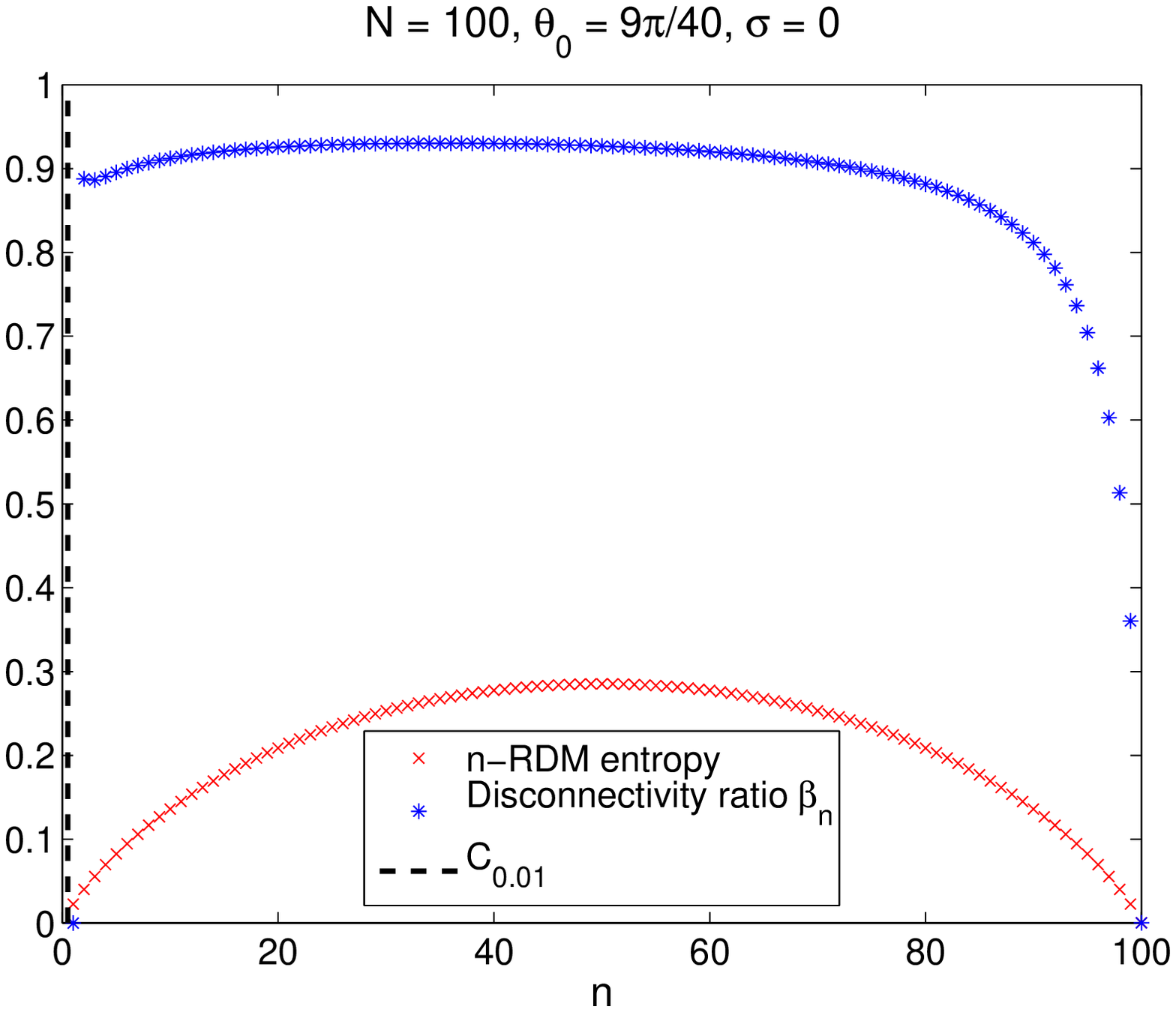} &
\includegraphics[width=0.5\textwidth]{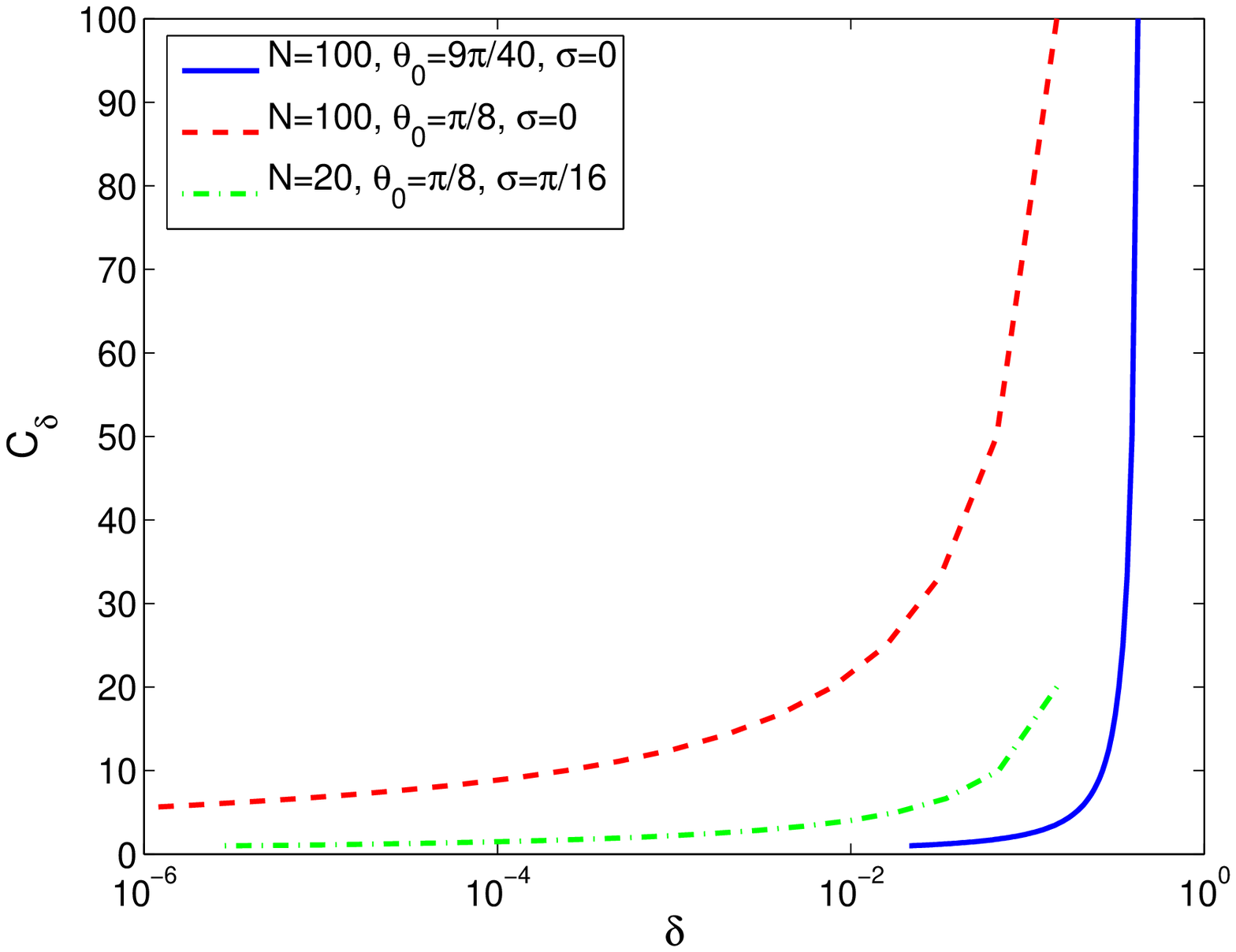}
\end{tabular}
\caption{(Color online.) $n$-RDM entropies $S_n$ and disconnectivity ratios $\beta_n$ for finite systems with $N$ bosons described by wavefunctions of the form of Eq.~(\ref{eq:BECIntegratedCat}) with three different sets of parameter values $\theta_0$ and $\sigma$. 
The disconnectivity $D$ is defined as the largest $n$ value for which $\beta_n \ll 1$ and is clearly equal to $N$ in all three cases shown here.  The vertical dashed line indicates cat size $C_{0.01}$ for precision $\delta = 0.01$. The lower right plot shows the effective cat size as a function of $\delta$ for the three cases.}
\label{fig:Disconnectivity}
\end{center}
\end{figure*}

We turn now to the entropy of $\tilde{\rho}^{(n)}$ for the  Gaussian cat states defined by Eqs.~(\ref{eq:BECIntegratedCat}) and~(\ref{eq:GaussianAmplitudeFunction}). This is plotted as a function of $n$ for various parameter combinations $\theta_0$ and $\sigma$ in Figure~\ref{fig:nRDMEntropies}, under the simplifying assumption that $N \gg n$ (since we restrict ourselves to this region, the drop in entropy as $n\rightarrow N$ cannot be seen). As expected from the above general arguments, when $\sigma = 0$ the entropies asymptotically approach $\ln 2$ as $n\rightarrow \infty$. This means that as we measure more and more particles, there exists a von~Neumann measurement with exactly two distinguishable and equally likely outcomes. In contrast, for $\sigma > 0$ the entropy of the $n$-RDM seems to grow without any upper bound, in an approximately logarithmic fashion. This means that, regardless of what kind of $n$-particle von~Neumann measurement we make, as $n \rightarrow \infty$ there will always be an ever increasing number of distinguishable outcomes. Our state is hence not just branching into a nice cat with two cleanly distinct branches, but instead developing a whole canopy! This canopy keeps growing with $n$. Hence it does not really make sense to view $\ket{\Psi}$ as any kind of two-branch or even a $d$-branch cat state in this situation. Instead, it is simply some more complicated kind of generic superposition state. 
(The zigzag-pattern for large values of $\sigma$ is caused by the factor of 
$(-1)^{n+k-l}$ in 
Eq.~(\ref{eq:S12S21sum}) in Appendix~\ref{app:RDMDerive}, which results in a different behaviour for even and odd values of $n$ when $\sigma \ne 0$, 
due to interference between contributions with a given $\theta$ in Eq.~ (\ref{eq:BECIntegratedCat}) for odd values of $n$.)

The von~Neumann entropy of the density matrix of the full state $\ket{\Psi}$ has previously been used to
 define a measure of cat size referred to as the disconnectivity $D$ by Leggett \cite{Leggett1980, Leggett2002}. To compute $D$, the entropy $S_n$ of the $n$-RDM is calculated for successively larger $n$. For each $n$ one also finds the minimum total entropy of any partition of the $n$ particles, i.e. $\min_{m} \left ( S_{m} + S_{n-m} \right )$, where the minimum is taken over all $m$ from $1$ to $n-1$. One then defines the ratio \footnote{Leggett denotes this ratio by $\delta_n$, but we have used $\beta_n$ to avoid confusion with the error threshold $\delta$ in our measure.}
\begin{equation}
\beta_n \equiv \frac{S_n}{\min_{1 \le m < n} \left ( S_{m} + S_{n-m} \right )},
\label{eq:DisconnectivityDeltan}
\end{equation}
and the disconnectivity of the system, $D$, is defined as the highest integer $n$ for which $\beta_n$ is smaller than some ``small'' fraction $\ll 1$ ($\beta_1$ is defined to be $0$). Thus $D = \mathrm{max}(n | \beta_n \ll 1)$. The motivation for this measure is that as long as $n$ is smaller than the total number of particles needed to observe perfectly the coherence of the 
joint state of all $N$ particles, the entropy $S_n$ will be nonzero since some information about the coherence is being neglected when $N-n$ particles are being traced out. Subdividing the system further will only neglect more information and increase the total entropy, so that $S_{m} + S_{n-m} > S_n$ and $\beta_n <1$. As $n$ approaches the number of particles sufficient to capture the full coherence of the system, $S_n$ and thus $\beta_n$ will approach zero. However, if $n$ can increase further beyond this point, then the denominator will \textit{also} vanish, and $\beta_n$ jumps again to $1$.
Thus the first value of $n$ at which all coherence is taken into account will be the largest number for which $\beta_n \ll 1$.
The term ``coherence'' is used here quite generally in the sense of correlations. If the system is made up of distinguishable particles and in a pure state, then these correlations will be equivalent to entanglement and the entropy $S_n$ of the $n$-RDMs is identical to the bipartite entanglement entropy between the $n$ particles included in the $n$-RDM and the $N-n$ particles being traced out~\cite{StocktonGeremiaDoherty2003}.
However, for indistinguishable particles, definition of entanglement must be made with care, since states with little ``useful'' entanglement can still look very entangled if one views single particles as good subsystems, due to the requirement that the total $N$-particle wavefunction be symmetrized or anti-symmetrized with respect to permutation of particles \cite{SchliemannCiracKus2001, Zanardi2002, Shi2003,WisemanBartlettVaccaro2003}. We will comment on these issues in more detail below.

There are both similarities and important differences between the disconnectivity and our measure of effective cat size. Both are based on considering how many particles must be measured to obtain a specific kind of information about the state or its components.
But while $C_{\delta}$  asks how many particles must be measured to \textit{differentiate} between the two branches composing the total state, the disconnectivity $D$ asks how many particles must be measured in order to observe all or nearly all correlations in the full quantum state. It also does not address whether or not the state is naturally divided into branches. These differences are reflected in very different numerical results. For the bosonic systems treated above, where we have assumed that $N$ is large and made approximations based on $n \ll N$ (see Appendix~\ref{app:RDMDerive} for a full description), 
explicit calculation for a range of $\sigma$ and $\theta_0$ values shows that $S_n$ increases monotonically with $n$ for the whole range treated (except for some minor oscillations between odd and even values of $n$), so that $\beta_n$ does not drop below $1/2$ until the assumption $n \ll N$ is no longer valid. This means that the disconnectivity must be of order $N$ for all parameter values $\sigma$ and $\theta_0$.  
In contrast, Figure~\ref{fig:RelativeCatSize} shows that for all values of $\sigma$ our measurement-based measure can give values of effective cat size $C_\delta$ much smaller than $N$, depending on the value of $\theta_0$.

In order to make a more direct comparison of $D$ with $C_\delta$, we have also calculated $S_n$ for $n$ from $1$ to $N$ for a finite value of $N$ and used this to evaluate the disconnectivity directly for some specific examples. We  use  $\theta_0 = 9\pi/40, \sigma = 0$ to study a system close to the full overlap situation ($\theta_0 = \pi/4$).  We use two examples at $\theta_0 = \pi/8$ ($\sigma = 0$, $N=100$ and $\sigma = \pi/16$, $N=20$) for study of an intermediate system and for analysis of the effect of nonzero spreading. The $n$-RDM entropies and disconnectivity ratios $\beta_n$, Eq.~(\ref{eq:DisconnectivityDeltan}), are plotted for these three cases in Figure~\ref{fig:Disconnectivity}.  The values of measurement-based cat size $C_{\delta}$ obtained for these parameters are superimposed as dashed vertical lines and the bottom right panel shows the sensitivity of $C_{\delta}$ to the precision $\delta$ for these three cases.
It is evident that for all three cases, $\beta_n$ is more or less constant at a value larger than one half and drops to a small fraction substantially smaller than this value only at $n\sim N$. Hence the disconnectivity $D$ is equal to or very close to $N$ in all cases. In contrast, our cat size measure based on distinguishability gives a cat size $C_{\delta}$ that is substantially less than $N$ for all three examples. With an error threshold $\delta = 0.01$, we obtain $C_{0.01} = N/5$ for $\theta_0 = \pi/8$ and $\sigma = 0$ or $\sigma = \pi/16$, and $C_{0.01} = 0$ for $\theta_0 = 9\pi/40$. Furthermore, the bottom right panel shows that in all three cases $C_{\delta} \ll N$ for all small $\delta$, so that our measure differs from disconnectivity for all reasonable error thresholds. 

This difference between disconnectivity and measurement-based cat size is not totally unexpected.
In order to observe perfectly all the correlations in the states of Eq.~\ref{eq:BECIntegratedCat}, one does indeed need to measure all or nearly all particles in the system, even when the branches are non-orthogonal, unless $\theta_0 = \pi/4$. However, it is clear that except when the branches are orthogonal, it is not possible to tell them apart with near certainty without measuring more than one particle and one hence obtains a reduced effective cat size. Only when we have a perfect cat with truly orthogonal branches, e.g., as in an ideal GHZ state, will the two measures agree.
For other states the two measures can be regarded as characterizing different aspects of the quantum correlations in a quantum state.

Another important aspect of disconnectivity can be seen by applying it not to cat-like states but to Fock states, i.e. states of the form $\ket{\Psi} \propto a^{\dag\, k} b^{\dag\, N-k} \ket{0}$.
For these states explicit calculation of the $n$-RDMs and their associated entropy $S_n$ shows that $D=N$ for all $k$ except $k=0$, where one obtains $D=1$ (see Appendix~\ref{app:FockStateD}). In contrast, since Fock states have no branches in the second-quantized formalism employed here, $n_{\min} > N$ and the measurement-based cat size measure gives a cat size $C_\delta = 0$ (see Section~\ref{sec:catsize}). They also have no entanglement when expressed in a second-quantized occupation-number basis. Thus it may seem puzzling that $D$ can be large.  However, we note that the disconnectivity relies on the entropies of the $n$-RDMs for its definition, and $S_n$ treats individual particles as the fundamental subsystems into which the system is divided and measures the correlation between them.  As noted in many recent papers, this is not appropriate if one is dealing with a system of indistinguishable particles, since the system can then appear to exhibit full $N$-particle entanglement simply due to the fact that the wavefunction has to be (anti-)symmetrized under exchange of particles. This fictitious entanglement, which has been referred to as ``fluffy bunny''-entanglement in the literature \cite{DunninghamRauBurnett2005}\cite{WisemanBartlettVaccaro2003} and which goes away if one treats only the modes as good subsystems instead of particles, is however necessarily present in the entropy of the $n$-RDM, $S_n$. The fluffy bunny entanglement contribution to disconnectivity is non-zero for all states other than those that can be written as Fock states with only a single occupied mode.  Consequently the disconnectivity of a system of indistinguishable particles will be large for all states that are not of this latter special kind, whether they are superposition states or not.  This suggests that one reason for the much larger values of $D$ than $C_{\delta}$ found here for the states of Eq.~(\ref{eq:BECIntegratedCat}) is inflation of the disconnectivity cattiness by fluffy bunny entanglement.  We note that redefining disconnectivity in terms of reduced density matrices of modes instead of particles, while possible in principle, will however be strongly dependent on the specific choice of modes. 
Nevertheless, a mode disconnectivity would be limited by the number of modes, and for a quantum condensate it is hence likely to also be substantially smaller than the total number of particles included in the description. 

\section{Application to cat states of BEC in a double well potential}
\label{sec:DoubleWellMonteCarlo}

\begin{figure*}
\begin{center}
\begin{tabular}{p{0.5\textwidth}p{0.5\textwidth}}
\includegraphics[width=0.5\textwidth]{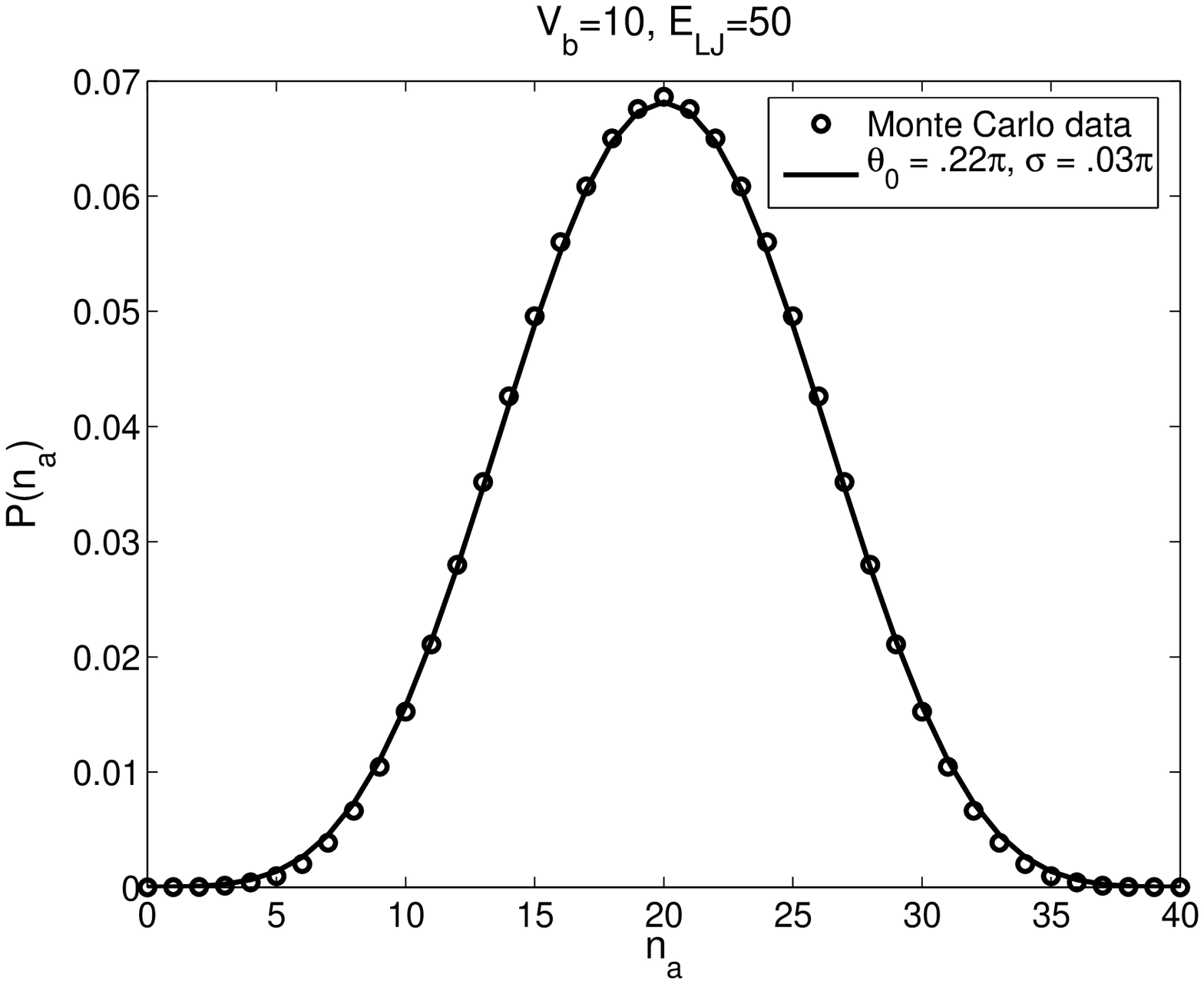} &
\includegraphics[width=0.5\textwidth]{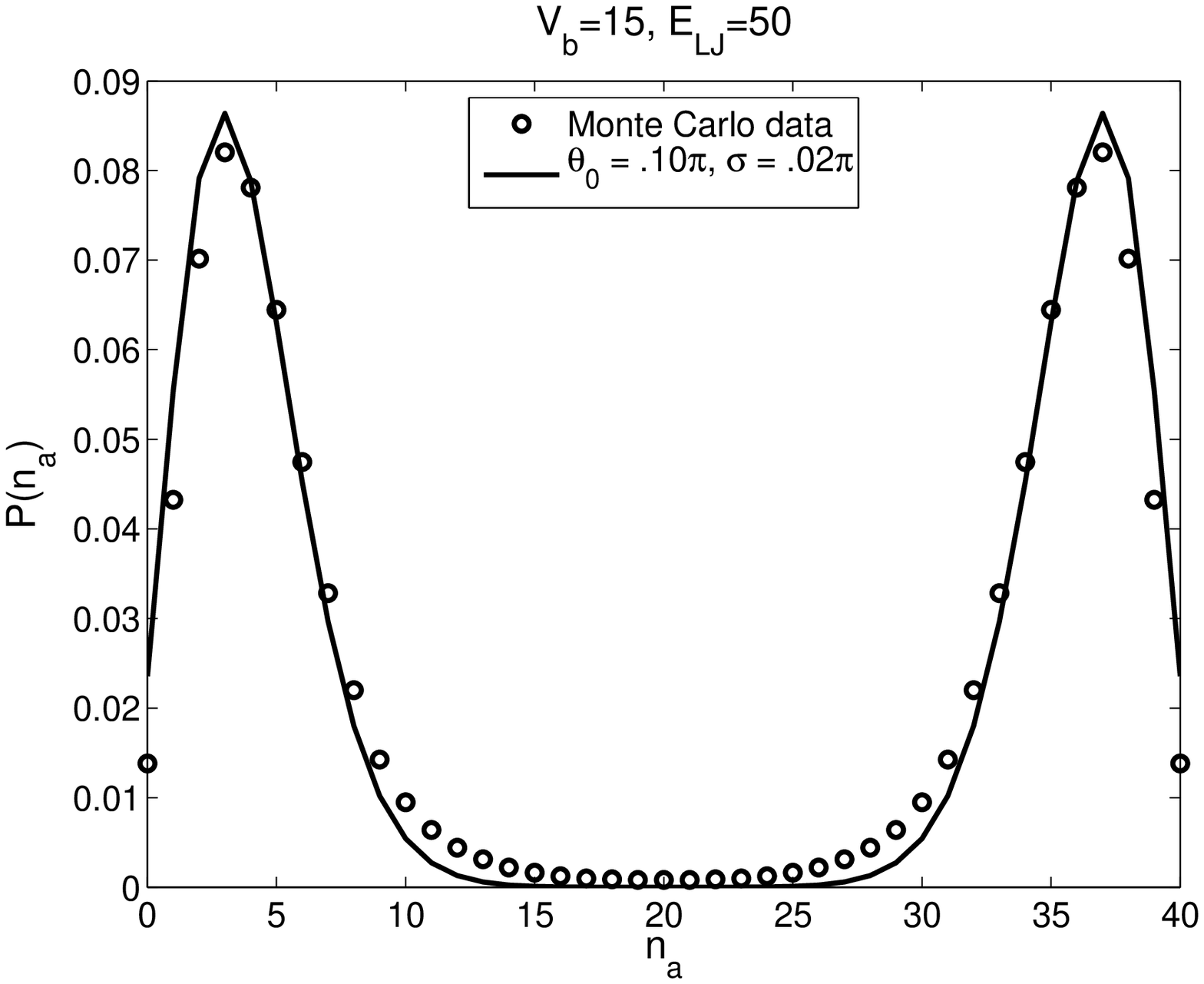} \\
\includegraphics[width=0.5\textwidth]{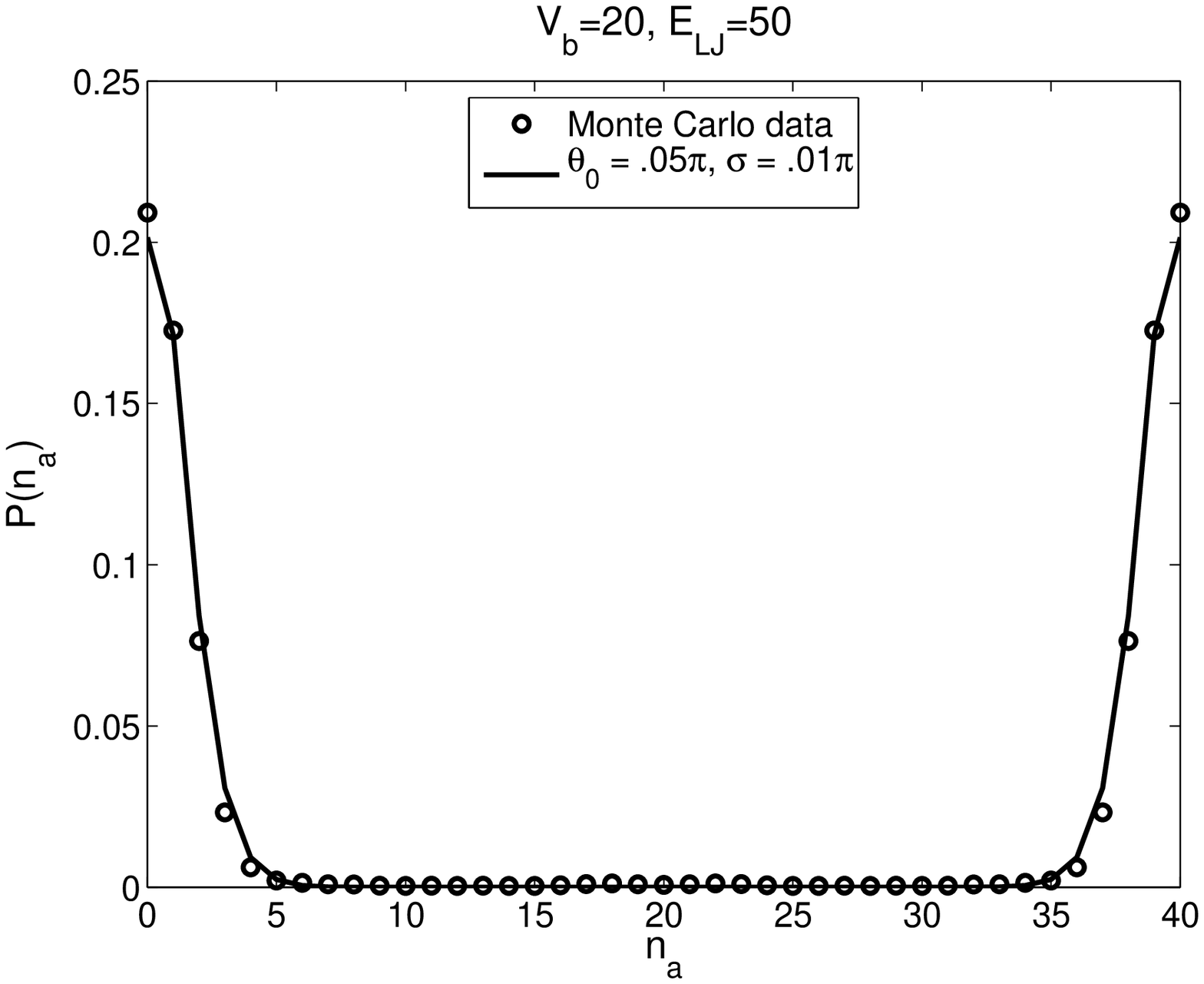} &
\includegraphics[width=0.5\textwidth]{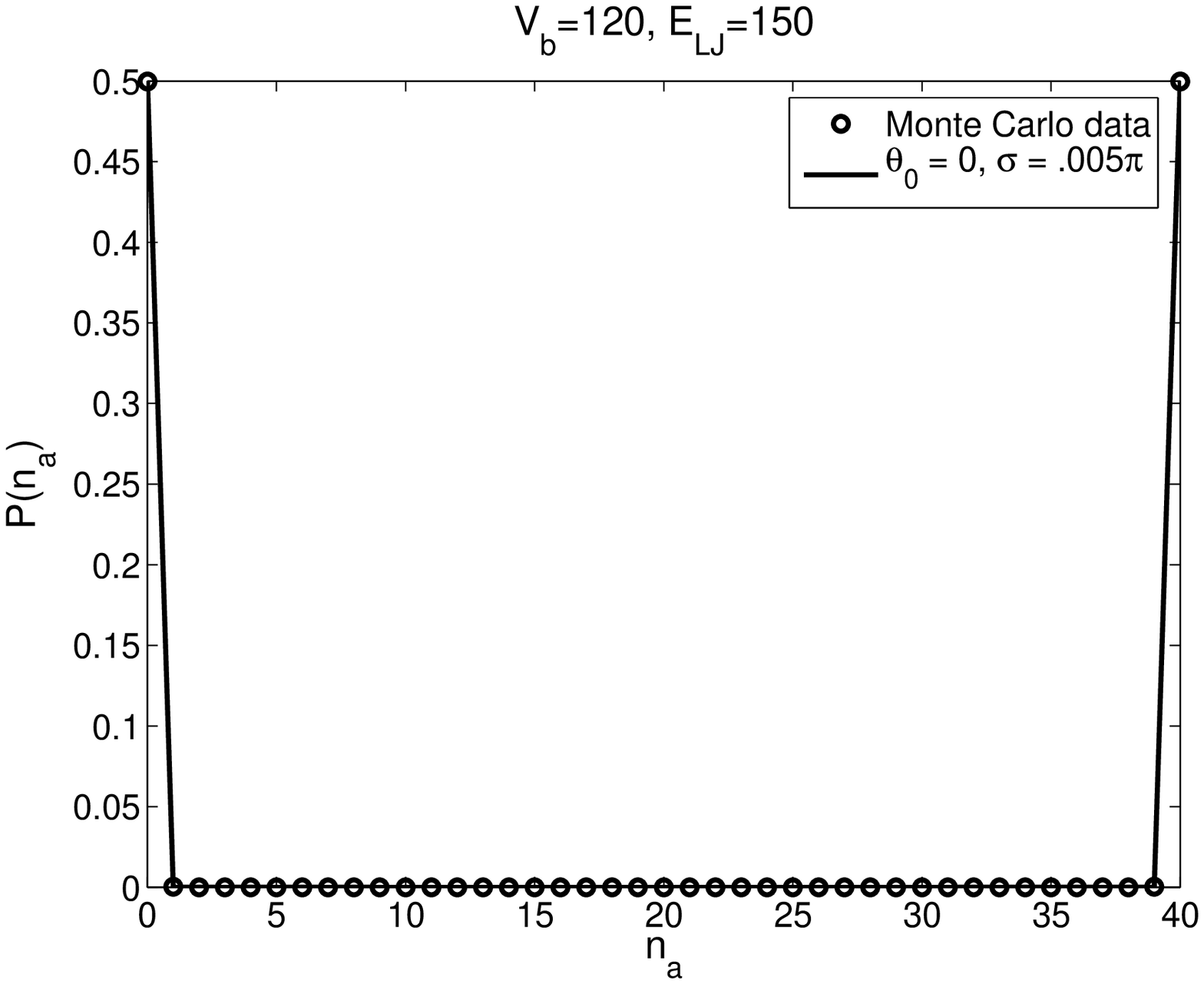}
\end{tabular}
\caption{Particle number distribution between two wells calculated from Monte Carlo simulation for bosons in a double well potential (open symbols) compared with best fit distributions given by Eq. (\ref{eq:BECIntegratedCat}) with a Gaussian spread function $f(\theta)$ (solid lines). $a_{\text{LJ}} = 0.15$ for all four cases.}
\label{fig:GaussAmplFit}
\end{center}
\end{figure*}

Finally, we apply our measure of cat size to a realistic system of bosons in a double well. We consider numerical results that have been obtained for bosons with attractive interactions in a spherically symmetric 3-dimensional harmonic trapping potential, which is split in two by a gaussian potential barrier in the $xy$-plane, forming a double well in the $z$-direction \cite{DuboisWhaley2006}. The 
numerical calculations were made using variational path integral Monte Carlo (VPI) \cite{CuervoRoyBoninsegni2005, SarsaSchmidtMagro2000} with 40 interacting bosonic atoms. The Hamiltonian used was\
\begin{multline}
H = \sum_{i} \left [ -\frac{1}{2} \nabla_i^2 + \frac{1}{2} r_i^2 + \frac{V_b}{\sqrt{4\pi\sigma^2}} \, e^{-\frac{z_i^2}{2\sigma^2}} \right ] \\ + \sum_{i\ne j} V_\text{int} \left ( \mathbf{r}_i - \mathbf{r}_j \right )
\end{multline}
where the sums run over the coordinates $\mathbf{r}_i$ of each of the 40 atoms and $V_b$ is a variable barrier height for the gaussian potential separating the two wells. Energies are given in units of $\hbar \omega /2$, where $\omega$ is the frequency of the ground state of the harmonic trapping potential, and lengths are given in units of $\sqrt{\hbar/m\omega}$. The two-particle interaction potential $V_{\text{int}}$ used here was a Lennard-Jones potential
\begin{equation}
V_{\text{int}}(r) = E_{\text{LJ}} \left [ \left ( \frac{a_{\text{LJ}}}{r} \right )^{12} - \left ( \frac{a_{\text{LJ}}}{r} \right )^6 \right ]
\end{equation}
with Lennard-Jones energy $E_{\text{LJ}}$ and length $a_{\text{LJ}}$.
The Lennard-Jones potential parameters $E_{LJ}$ and $a_{LJ}$ determine the scattering length $a$~\cite{FlambaumGribakinHarabati1999}. It thus provides a model potential that allows us to design a computationally efficient sampling scheme for a given scattering length, $a$~\cite{DuboisWhaley2006}.  Formation of cat states require a negative value of $a$.  For a realistic cold atom systems with attractive effective interactions such as $^7$Li ($a = -14.5$ \AA~\cite{AbrahamMcAlexanderGerton1997}), we find that stable cat states can be formed with $\sim 1000$ atoms in a trap of linear dimension $a_{ho} = \hbar/m\omega = 13,000$ \AA, using suitable  values of Lennard-Jones parameters.

\begin{table}
\begin{center}
\begin{tabular*}{0.45\textwidth}{@{\extracolsep{\fill}}|c|c|c|c|c|c|}
\hline $V_b$ & $E_{\text{LJ}}$ & Best fit $\theta_0$ & Best fit $\sigma$ & $C_{0.01}$ & $C_{10^{-4}}$ \\ \hline
$10$ & $50$ & $.22\pi$ & $.030\pi$ & $0$ & $0$ \\
$15$ & $50$ & $.10\pi$ & $.020\pi$ & $10$ & $4$ \\
$20$ & $50$ & $.05\pi$ & $.010\pi$ & $20$ & $10$ \\
$120$ & $150$ & $0$ & $.005\pi$ & $40$ & $40$ \\
\hline
\end{tabular*}
\caption{Best fit of $\theta_0$, $\sigma$, and effective cat sizes $C_{0.01}$ at $\delta = 0.01$ and
$C_{10^{-4}}$ at $\delta = 10^{-4}$,
for four numerically calculated distributions of bosons in a double-well potential. $a_{\text{LJ}}=0.15$ in all cases.}
\label{tbl:Fits}
\end{center}
\end{table}

To compare with our model states in Eq.~(\ref{eq:BECIntegratedCat}), the numerical data was used to find the probability distribution $P(n$) for finding $n$ of the $N=40$ particles on one side of the double-well. This was done for three cases with $V_b = 10, 15, 20$ and with $E_{\text{LJ}} = 50$, $a_{\text{LJ}} = .15$ in all three cases, and for one case with $V_b = 120$, $E_{\text{LJ}} = 150$, and $a_{\text{LJ}} = .15$ (the last choice of extra high potential barrier and strong attractive interaction was made to get as close to a maximal cat state as possible). We then fit the probability distribution $P(n_a)$ for the number of particles in mode $a$ calculated from the states in Eq.~(\ref{eq:BECIntegratedCat}), to the numerically calculated distributions in each case by varying $\theta_0$ and $\sigma$ to obtain the smallest possible difference between the two distributions in the least mean square sense. The fitting had a resolution of $0.10\pi$ in $\theta_0$ and $.005\pi$ in $\sigma$.
The resulting best fit values for each case are shown in Table~\ref{tbl:Fits}, along with the effective cat sizes $C_{0.01}$ for $\delta = 0.01$ and $C_{10^{-4}}$ for $\delta = 10^{-4}$ calculated using the states of Eq.~(\ref{eq:BECIntegratedCat}) with the fitted values of $\theta_0$ and $\sigma$ (the numerical precision in the calculations do not warrant smaller values of $\delta$). The corresponding fitted number distributions are compared to the VPI distributions in 
Figure~\ref{fig:GaussAmplFit}, showing a very good fit for the cases studied here. Note that this does not imply that our states give the correct phases between the superposed states, since we are only fitting to the number distribution. However, given that Eq.~(\ref{eq:BECIntegratedCat}) with $\sigma = 0$ gives the exact ground state in the mean-field limit \cite{CiracLewensteinMolmer1998}, it is reasonable to expect that Eq.~(\ref{eq:BECIntegratedCat}) constitutes a good approximation to the true states. Our comparison with the distributions calculated from VPI Monte Carlo supports this expectation and also implies that the probability distributions (but not necessarily the amplitudes~\cite{DuboisWhaley2006}) can be accurately described by a two-mode approximation.

Table~\ref{tbl:Fits} shows that for the lowest barrier height $V_b=10$ we do not really get a cat state at all, since the low barrier height results in large tunneling, which allows the particles to overcome their attractive interactions and distribute themselves almost binomially between the two wells. The best fit value of $\theta_0$ ($0.22\pi$), is less than one $\sigma$ away from the complete-overlap value $\pi/4$,
 and the effective cat size is correspondingly zero since the branches are strongly overlapping. As the barrier height $V_b$ is increased for a given attraction strength $E_{\text{LJ}}$, the tunneling rate decreases, and it becomes more favorable for all particles to sit in one well. However, since the tunneling amplitude is still finite, the lowest-energy state is not a Fock state but rather a superposition state of nearly all particles being in either one well or the other, {\it i.e.}, a cat state. Thus, $C_{\delta}$ increases with $V_b$. 
In the most extreme example here, $V_b=120, E_{LJ}=150$, the tunneling amplitude is extremely small and the branches have negligible overlap, resulting in an ideal cat state $C_{\delta}=40$ for $N=40$.
As expected, we see that $C_{\delta}$ does depend on the value of the precision $\delta$, becoming smaller as $\delta$ decreases. We also see that the decrease in cat size is greater for larger $\sigma$ values, while for the most ``catty'' case ($V_b=120$ and $E_{\text{LJ}} = 150$), where $\sigma$ is practically zero, $C_{\delta}$ is not affected at all by reducing $\delta$ from $10^{-2}$ to $10^{-4}$.

We also calculated the disconnectivity $D$ for these states and find that $D=N=40$ in all four cases. This may appear initially somewhat surprising, especially for the case of $V_b = 10$ (top left panel in Figure~\ref{fig:GaussAmplFit}, since in that state the branches are almost completely overlapping, and resemble a binomially distributed state more than a cat state. However, even in this case, since the distribution is not exactly binomial, there must be some entanglement between the particles.  Furthermore, all $N$ particles must be involved in this entanglement since they are indistinguishable. As discussed in Section~\ref{sec:entropies_disconnectivity}, this coherence between all particles leads to a large value for $D$, even though the state cannot be reasonably called a cat state in any way.

\section{Conclusions and future work}

We have presented a measure of the effective size of superposition states in general quantum systems,
{\it i.e.}, the number of effective subsystems that can describe the superposition, that is based on how well measurements can distinguish between the different branches of the state. 
Our measure does in general require one to consider coherent multi-particle measurements, 
although we find that for the special class of states considered in \cite{DurSimonCirac2002}, a procedure using only single-particle measurements can be useful.
The resulting "cat size" measure is dependent on the precision to which the branches are to be distinguished. 
Application of this measurement-based measure to generalized superpositions states of bosons in a two-mode system predicts cat sizes much smaller than what is predicted from the earlier measure of disconnectivity that was proposed in~\cite{Leggett1980}. 
Analysis of disconnectivity for specific examples showed that for indistinguishable particles this quantity is large for a much wider variety of states than superposition states, including single-branch Fock states, due to the inclusion of particle correlations induced by \hbox{(anti-)}symmetrization.

We expect that the new measure will be useful for comparing the effective size of superposition states in different kinds of physical systems, including those with macroscopic numbers of constituents.  We have shown that the generalized superposition states studied here can be fit to realistic numerical simulations of bosons in a 3D double-well trapping potential, and have analyzed the cattiness of superposition states of these interacting bosons as a function of their interaction strength and of the barrier height.
Future directions include applying our measure to more complicated systems that have been realized experimentally, in particular to the experiments with superconducting loops reported in \cite{WalHaarWilhelm2000} and \cite{FriedmanPatelChen2000}. In a very recent paper \cite{MarquardtAbelDelft2006}, a different cat size measure was defined and applied to the three-Josephson junction circuit reported in \cite{WalHaarWilhelm2000}, and the cat size according to that measure found to be extremely small (of order 1).  It would thus be of great interest to evaluate the new measurement-based measure of cat size for superpositions of superconducting loops.

\acknowledgments{
The authors thank  J.~von~Delft, F.~Wilhelm, F.~Marquardt and A.~J.~Leggett for useful discussions. This research effort was sponsored by the Defense Advanced Research project Agency (DARPA), 
the Air Force Laboratory, Air Force Material Command, USAF, under
contract No. F30602-01-2-0524, and in part by the National Science Foundation through the San Diego Supercomputer Center under grant UCB232 using Datastar. J.~I.~Korsbakken also acknowledges support from the Research Council of Norway.
}

\appendix

\section{Calculation of $n$-particle reduced density matrices}
\label{app:RDMDerive}

Inner products between the states $\ket{\phi_{1,2}^{(N)}(\theta)}$ are computed in the $c, d$ basis using standard methods, giving
\begin{widetext}

\begin{equation}
\begin{split}
\braket{\phi_A^{(N)}(\theta)}{\phi_A^{(N)}(\theta')} &= \frac{1}{2^N} \bra{0} \left ( e^{-i\theta} c - ie^{i\theta} d \right )^N \left ( e^{i\theta'} c^{\dag} + ie^{-i\theta'} d^{\dag} \right )^N \ket{0} \\
&= \frac{1}{2^N} N \left ( e^{i(\theta - \theta')} + e^{-i(\theta - \theta')} \right ) \bra{0} \left ( e^{-i\theta} c - i e^{i\theta} d \right )^{N-1} \left ( e^{i\theta'} c^{\dag} + ie^{-i\theta'} d^{\dag} \right )^{N-1} \ket{0} \\
&\equiv N \cos \left ( \theta - \theta' \right ) \braket{\phi_1^{(N-1)}(\theta)}{\phi_1^{(N-1)}(\theta')} \\
&= N(N-1) cos^2 \left (\theta - \theta' \right ) \braket{\phi_1^{(N-2)}(\theta)}{\phi_1^{(N-2)}(\theta')} \\
& \dots \\
& = N! \, \cos^N \left ( \theta - \theta' \right ) \simeq N! \, \sqrt{\frac{2\pi}{N}} \, \delta \left ( \theta - \theta' \right )
\end{split}
\label{eq:phiAphiA}
\end{equation}

and similarly

\begin{align}
\braket{\phi^{(N)}_B(\theta)}{\phi^{(N)}_B(\theta')} &= N! \, \cos^N \left ( \theta - \theta' \right ) \simeq N! \, \sqrt{\frac{2\pi}{N}} \, \delta \left ( \theta - \theta' \right )
\label{eq:phiBphiB} \\
\braket{\phi^{(N)}_A(\theta)}{\phi^{(N)}_B(\theta')} &= N! \, \sin^N \left ( \theta + \theta' \right ) \simeq N! \, \sqrt{\frac{2\pi}{N}} \, \left [ \delta \left ( \frac{\pi}{2} - \theta - \theta ' \right ) + \delta \left ( -\frac{\pi}{2} - \theta - \theta' \right ) \right ].
\label{eq:phiAphiB}
\end{align}

\end{widetext}

\noindent The $\delta$-function approximations are valid in the limit of large $N$. We have assumed that $\theta + \theta'$ is bounded to lie between $\pm \pi/2$.

Defining
\begin{equation}
(\tilde{\rho}_{\alpha\beta}^{(n)})^k_l \equiv \sqrt{{n \choose k}{n \choose l}} \, \bra{\Psi_{\alpha}^{(N)}} c^{\dag \, k} d^{\dag \, n-k} \, c^{l} d^{n-l} \ket{\Psi_{\beta}^{(N)}}
\end{equation}
with $\alpha\beta = AA$, $BB$, $AB$ or $BA$ ($\tilde{\rho}_{AA}$ and $\tilde{\rho}_{BB}$ correspond to $\tilde{\rho}_A^{(n)}$ and $\tilde{\rho}_B^{(n)}$ as defined in \ref{subsec:bosonic_effectivecatsizes}), and using the action of the operators $c, d$ on the branches 
$\ket{\phi_1^{(N)}(\theta)}$ and $\ket{\phi_2^{(N)}(\theta)}$ leads to:

\begin{widetext}
\begin{align}
\left ( \tilde{\rho}_{AA}^{(n)} \right )^k_l &\propto \frac{i^{k-l}}{2^n} \sqrt{{n\choose k}{n\choose l}} \int_{\CLMZLowerlim}^{\CLMZUpperlim} \diff \theta \, \diff \theta' \, f(\theta)^* f(\theta') \, e^{-2i(k\theta - l\theta')+in(\theta-\theta')}\cos^{N-n}(\theta-\theta') \nonumber \\
&\simeq \frac{i^{k-l}}{2^n} \sqrt{{n\choose k}{n\choose l}} \int_{\CLMZLowerlim}^{\CLMZUpperlim} \diff \theta \, e^{-2i(k-l)\theta} \left | f(\theta) \right |^2 
\label{eq:rhoAAResult} \\
\left ( \tilde{\rho}_{BB}^{(n)} \right )^k_l &\propto \frac{i^{-(k-l)}}{2^n} \sqrt{{n\choose k}{n\choose l}} \int_{\CLMZLowerlim}^{\CLMZUpperlim} \diff \theta\, \diff \theta' \, f(\theta)^* f(\theta') \, e^{2i(k\theta-l\theta')-in(\theta-\theta')} \cos^{N-n}(\theta - \theta') \nonumber \\
&\simeq \frac{i^{-(k-l)}}{2^n} \sqrt{{n\choose k}{n\choose l}} \int_{\CLMZLowerlim}^{\CLMZUpperlim} \diff \theta \, e^{2i(k-l)\theta}\left | f(\theta) \right |^2
\label{eq:rhoBBResult} \\
\left ( \tilde{\rho}_{AB}^{(n)} \right )^k_l &\propto \frac{i^{k+l-n}}{2^n} \sqrt{{n\choose k}{n\choose l}} \int_{\CLMZLowerlim}^{\CLMZUpperlim} \diff \theta \, \diff \theta' \, f(\theta)^* f(\theta') \, e^{-2i(k\theta+l\theta')+in(\theta+\theta')} \sin^{N-n}(\theta+\theta') \nonumber \\
&\simeq \frac{i^{k-l}}{2^n} \sqrt{{n\choose k}{n\choose l}} \int_{\CLMZLowerlim}^{\CLMZUpperlim} \diff \theta \, e^{-2i (k-l)\theta} \left [ f(\theta)^* f(\pi/2-\theta) + (-1)^n f(\theta)^* f(-\pi/2 - \theta) \right ]
\label{eq:rhoABResult} \\
\left ( \tilde{\rho}_{BA}^{(n)} \right )^k_l &= \left [ \left ( \tilde{\rho}_{AB}^{(n)} \right )^l_k \right ]^* \propto \frac{i^{n-k-l}}{2^n} \sqrt{{n\choose k}{n\choose l}} \int_{\CLMZLowerlim}^{\CLMZUpperlim} \diff \theta \, \diff \theta' \, f(\theta) f(\theta')^* \, e^{2i(l\theta+k\theta')-in(\theta+\theta')} \sin^{N-n}(\theta+\theta') \nonumber \\
&\simeq \frac{i^{k-l}}{2^n} \sqrt{{n\choose k}{n\choose l}} \int_{\CLMZLowerlim}^{\CLMZUpperlim} \diff \theta \, e^{-2i(k-l)\theta} \left [ f(\theta) f(\pi/2-\theta)^* + (-1)^n f(\theta)^* f(-\pi/2-\theta) \right ]
\label{eq:rhoBAResult}
\end{align}
\end{widetext}
where in the last steps we have made use of the above delta function approximation.

Using the gaussian form in Eq.~(\ref{eq:GaussianAmplitudeFunction}) for the amplitude spreading function, results in the following analytic forms for the $n$-RDM matrix elements:

\begin{align}
\left ( \tilde{\rho}_{AA}^{(n)} \right )^k_l &= \frac{i^{k-l}}{2^n} \sqrt{{n\choose k}{n\choose l}} \, e^{-2i(k-l)\theta_0} e^{-2(k-l)^2\sigma^2} \\
\left ( \tilde{\rho}_{BB}^{(n)} \right )^k_l &= \frac{i^{-(k-l)}}{2^n} \sqrt{{n\choose k}{n\choose l}} \, e^{2i(k-l)\theta_0} e^{-2(k-l)^2\sigma^2}
\end{align}

\noindent The $n$-RDM of the full state $\ket{\Psi}$, which we use for calculating entropies in \ref{sec:entropies_disconnectivity}, requires also the sum of $\tilde{\rho}_{AB}^{(n)}$ and $\tilde{\rho}_{BA}^{(n)}$ which is given by

\begin{multline}
\left ( \tilde{\rho}_{AB}^{(n)} + \tilde{\rho}_{BA}^{(n)} \right )^k_l = \frac{2}{2^n} \sqrt{{n\choose k}{n\choose l}} \, e^{-2(k-l)^2 \sigma^2} \\
\times \left [ e^{-\frac{(\theta_0 - \pi/4)^2}{2\sigma^2}} + (-1)^{n+k-l} e^{-\frac{(\theta_0 + \pi/4)^2}{2\sigma^2}} \right ].
\label{eq:S12S21sum}
\end{multline}
The traces of the two first matrices are already equal to $1$, so no further normalization is necessary. The trace of the matrix defined in Eq.~(\ref{eq:S12S21sum}) is given by 
\begin{multline}
\sum_{k=0}^n \left ( \tilde{\rho}_{AB}^{(n)} + \tilde{\rho}_{BA}^{(n)} \right )^k_k \\
=  2 \left ( e^{-\frac{(\theta_0-\pi/4)^2}{2\sigma^2}} + \left ( -1 \right )^n e^{-\frac{(\theta_0+\pi/4)^2}{2\sigma^2}} \right )
\end{multline}
so that the final form of the symmetrized $n$-RDM $\tilde{\rho}^{(n)}$, properly normalized, is
\begin{widetext}
\begin{equation}
\begin{split}
\left ( \tilde{\rho}^{(n)} \right )^k_l &\equiv \frac{\left ( \tilde{\rho}^{(n)}_{AA} + \tilde{\rho}^{(n)}_{BB} + \tilde{\rho}^{(n)}_{AB}+\tilde{\rho}^{(n)}_{BA} \right )^k_l}{\trace \left ( \tilde{\rho}^{(n)}_{AA} + \tilde{\rho}^{(n)}_{BB} + \tilde{\rho}^{(n)}_{AB} + \tilde{\rho}^{(n)}_{BA} \right )} \\
&= \sqrt{{n \choose k}{n \choose l}} \, \frac{i^{k-l} e^{-2i(k-l)\theta_0} + i^{-(k-l)} e^{2i(k-l)\theta_0} + 2 \left [ E_{-} + (-1)^{n+k-l} E_{+} \right ]}{2^n \left [ 2 + 2 \left ( E_{-} + (-1)^{n+k-l} E_{+} \right ) \right ]} \, e^{-2(k-l)^2 \sigma^2}
\end{split}
\end{equation}
\end{widetext}
where $E_{\pm} \equiv \exp \left [ - \left (\theta_0 \pm \pi/4 \right )^2/2\sigma^2 \right ]$.

\section{Derivation of $q^{(k+1)}_{\gamma} = P_k$ for $\gamma = A$ or $B$}
\label{app:PosteriorProbs}

To show that one of the prior probabilities $q^{(k+1)}_{A,B}$ of branch A or B before performing the $(k+1)$'th measurement in Section~\ref{subsec:bosonic_mx1} will be equal to the success probability $P_k$ of identifying the correct branch in the $k$'th measurement, first note that using Bayes' theorem and the definition of conditional probabilities, the success probability $P_k$ as given by Eq.~(\ref{eq:Pk}) can equivalently be written as
\begin{equation}
\begin{split}
P_k &= P(A|E^{(k)}_A) P(E^{(k)}_A) + P(B|E^{(k)}_B) P(E^{(k)}_B)
\label{eq:AltPk}
\end{split}
\end{equation}
where $P(\gamma | E^{(k)}_{\delta})$ is the posterior probability that the state is $\ket{\Psi_{\gamma}}$ \textit{given} that the measurement on the $k$'th particle gave the outcome $E^{(k)}_{\delta}$, and $P(E^{(k)}_{\gamma})$ is the total probability that the measurement gives the outcome $E^{(k)}_{\gamma}$, \textit{irrespective} of what the state is. Second, note that $P(A|E^{(k)}_A)$ and $P(B|E^{(k)}_B)$ are in fact equal. This follows from
\begin{equation}
\begin{split}
P(A|E^{(k)}_A) &= \frac{P(E^{(k)}_A|A) q^{(k)}_A}{P(E^{(k)}_A)} \\
&= \frac{P(E^{(k)}_A|A) q^{(k)}_A}{q^{(k)}_A P(E^{(k)}_A|A) + q^{(k)}_B P(E^{(k)}_A| B)} \\
&= \frac{1}{1 + \frac{q^{(k)}_B P(E^{(k)}_A|B)}{q^{(k)}_A P(E^{(k)}_A|A)}}
\end{split}
\end{equation}
and similarly
\begin{equation}
\begin{split}
P(B|E^{(k)}_B) &= \frac{P(E^{(k)}_B|B) q^{(k)}_B}{q^{(k)}_A P(E^{(k)}_B|A) + q^{(k)}_B P(E^{(k)}_B|B)} \\
&= \frac{1}{1+ \frac{q^{(k)}_A P(E^{(k)}_B|A)}{q^{(k)}_B P(E^{(k)}_B|B)}} \, .
\end{split}
\end{equation}
These will be equal iff
\begin{equation}
\frac{q^{(k)}_B P(E^{(k)}_A|B)}{q^{(k)}_A P(E^{(k)}_A|A)} = \frac{q^{(k)}_A P(E^{(k)}_B|A)}{q^{(k)}_B P(E^{(k)}_B|B)} \, .
\label{eq:PosteriorsEqualCondition}
\end{equation}
After a good deal of algebra, using Eqs.~(\ref{eq:PEAGivenA})--(\ref{eq:PEBGivenB}), the fact that $q^{(k)}_A + q^{(k)}_B = 1$, and moving factors between the two sides of Eq.~(\ref{eq:PosteriorsEqualCondition}), both sides can be reduced to
\begin{equation}
\left ( q^{(k)}_A q^{(k)}_B \right )^2 \cos^2 2\theta_k \left ( 1 - \cos^2 2\theta_k \right ) \, ,
\end{equation}
proving that indeed $P(A|E^{(k)}_A) = P(B|E^{(k)}_B)$. Finally, since the measurement on particle $k$ must give either the outcome $E^{(k)}_A$ or $E^{(k)}_B$, we have $P(E^{(k)}_A) + P(E^{(k)}_B) = 1$, so that Eq.~(\ref{eq:AltPk}) reduces to
\begin{equation}
P_k = P(A|E^{(k)}_A) = P(B|E^{(k)}_B)\, ,
\end{equation}
which is what we wanted to show.

\section{Disconnectivity of Fock states}
\label{app:FockStateD}

In this appendix we show that the disconnectivity, $D$, determined by Eq.~(\ref{eq:DisconnectivityDeltan}) is equal to the total particle number $N$ for all Fock states that have more than one mode with non-zero occupation number.

A Fock state in a second-quantized system with $d$ modes, occupation numbers $\mathbf{n} \equiv (n_1, n_2, \ldots, n_d)$ and a total of $N$ particles has the form
\begin{equation}
\ket{\mathbf{n}} \, \equiv \, \prod_{k=1}^d \frac{(a_k^{\dag})^{n_k}}{\sqrt{n_k !}} \, \ket{0}
\label{eq:FockState}
\end{equation}
with $\sum_k n_k = N$. We assume here that the particles are bosons, although this does not affect our final conclusion. We then define a symmetrized $n$-RDM $\tilde{\rho}^{(n)}$ by generalizing Eqs.~(\ref{eq:SymmetrizedBosonicRDM}) and~(\ref{eq:BosonicRDM}). For this we use $\mathbf{p} \equiv (p_1, p_2, \ldots, p_d)$ and $\mathbf{q} \equiv (q_1, q_2, \ldots, q_d)$ as upper and lower indices, representing the number of creation and annihilation operators, respectively,
\begin{multline}
\left ( \tilde{\rho}^{(n)} \right )^{\mathbf{p}}_{\mathbf{q}} \equiv \frac{(N-n)!}{N!} \frac{n!}{\sqrt{\prod_k p_k ! \, q_k!}} \\ \bra{\Psi} (a_d^{\dag})^{p_1} \cdots (a_1^{\dag})^{p_d} a_1^{q_1} \cdots a_d^{q_d} \ket{\Psi},
\label{eq:ManyModenRDM}
\end{multline}
subject to the constraint that $\sum_k p_k = \sum_k q_k = n$.
For a Fock state 
Eq.~(\ref{eq:ManyModenRDM}) is non-zero only for $\mathbf{p}=\mathbf{q}$, i.e. the $n$-RDM is diagonal. Furthermore, we must have $p_k, q_k \leq n_k$ for a given matrix element not to vanish. For the case $N=n$, the only non-zero matrix element is then $\mathbf{p} = \mathbf{q} = \mathbf{n}$, {\it i.e.} the $N$-RDM $\tilde{\rho}^{(N)}$ has only a single matrix element equal to $1$ on the diagonal and the rest are equal to zero. Hence the entropy is $S_N = 0$. On the other hand, if $n<N$ and if there is more than one $n_k > 0$, there will be at least two different $\mathbf{p} = \mathbf{q}$ for which $(\tilde{\rho}^{(n)})^{\mathbf{p}}_{\mathbf{p}} \neq 0$, so that $\tilde{\rho}^{(n)}$ must have more than one non-zero eigenvalue. Therefore $S_n > 0$ for all $n<N$. This implies that the numerator of $\beta_N$ in Eq.~(\ref{eq:DisconnectivityDeltan}) vanishes while the denominator does not. Hence $\beta_N = 0$, so that $n = N$ is the largest $n$ for which $\beta_n \ll 1$, and consequently the disconnectivity is $D = N$, provided that there is more than one mode with non-zero occupation number.
If only one mode is occupied, $\tilde{\rho}^{(n)}$ has only a single non-zero eigenvalue (equal to $1$) for \textit{all} $n$, and therefore $\beta_n = 1$ for all $n>1$. Since  $\beta_1 = 0$ by definition, we therefore have $D=1$ for a Fock state in which only a single mode is occupied.

\bibliographystyle{apsrev}
\bibliography{catstate}

\end{document}